\pdfoutput=1
\documentclass[journal]{IEEEtran}

\usepackage{graphicx,subcaption,xcolor,setspace,amsthm,amsmath,cite,booktabs,epstopdf,dblfloatfix,blindtext,amsthm}
\usepackage[hyphens]{url}
\usepackage{bm}
\usepackage{algorithmic}
\usepackage{upgreek,acro}
\usepackage{bbm}
\usepackage[ruled,norelsize]{algorithm2e}
\usepackage[binary-units,detect-weight=true, detect-family=true]{siunitx}
\DeclareSIUnit \bitspersecond {bps}
\usepackage{amsmath,amssymb}

\newcommand{\bbC}{{\mathbb{C}}}

\newcommand{\bbE}{{\mathbb{E}}}

\newcommand{\bbR}{{\mathbb{R}}}

\newcommand{\ba}{{\mathbf{a}}}

\newcommand{\bd}{{\mathbf{d}}}

\newcommand{\br}{{\mathbf{r}}}

\newcommand{\bu}{{\mathbf{u}}}
\newcommand{\bv}{{\mathbf{v}}}
\newcommand{\bw}{{\mathbf{w}}}
\newcommand{\bx}{{\mathbf{x}}}

\newcommand{\bzero}{{\mathbf{0}}}

\newcommand{\bC}{{\mathbf{C}}}

\newcommand{\bF}{{\mathbf{F}}}

\newcommand{\bH}{{\mathbf{H}}}
\newcommand{\bI}{{\mathbf{I}}}

\newcommand{\bQ}{{\mathbf{Q}}}
\newcommand{\bR}{{\mathbf{R}}}

\newcommand{\bT}{{\mathbf{T}}}

\newcommand{\bW}{{\mathbf{W}}}
\newcommand{\bX}{{\mathbf{X}}}
\newcommand{\bY}{{\mathbf{Y}}}

\newcommand{\rmc}{{\mathrm{c}}}

\newcommand{\rmm}{{\mathrm{m}}}

\newcommand{\rmr}{{\mathrm{r}}}
\newcommand{\rms}{{\mathrm{s}}}

\newcommand{\rmB}{{\mathrm{B}}}

\newcommand{\rmF}{{\mathrm{F}}}

\newcommand{\rmR}{{\mathrm{R}}}
\newcommand{\rmS}{{\mathrm{S}}}

\newcommand{\rmU}{{\mathrm{U}}}
\newcommand{\rmV}{{\mathrm{V}}}

\newcommand{\cC}{\mathcal{C}}

\newcommand{\cN}{\mathcal{N}}

\newcommand{\cU}{\mathcal{U}}

\newcommand{\bsfn}{\boldsymbol{\mathsf{n}}}

\newcommand{\bsfs}{\boldsymbol{\mathsf{s}}}

\newcommand{\bsfx}{\boldsymbol{\mathsf{x}}}
\newcommand{\bsfy}{\boldsymbol{\mathsf{y}}}

\newcommand{\bsfF}{\boldsymbol{\mathsf{F}}}

\newcommand{\bsfH}{\boldsymbol{\mathsf{H}}}

\newcommand{\bsfR}{\boldsymbol{\mathsf{R}}}

\newcommand{\bsfW}{\boldsymbol{\mathsf{W}}}

\newcommand{\transp}{{\sf T}}
\newcommand{\compj}{{\rm j}}

\makeatletter
\def\munderbar#1{\underline{\sbox\tw@{$#1$}\dp\tw@\z@\box\tw@}}
\makeatother
\newcommand{\dBm}{\SI{}{\decibel}\rmm}

\DeclareAcronym{3GPP}{
	short=3GPP,
	long=3rd Generation Partnership Project
}
\DeclareAcronym{ADC}{
	short=ADC,
	long=analog-to-digital converter
}
\DeclareAcronym{AMP}{
	short=AMP,
	long=approximate message passing
}
\DeclareAcronym{AoA}{
	short=AoA,
	long=angle-of-arrival
}
\DeclareAcronym{AoD}{
	short=AoD,
	long=angle-of-departure
}
\DeclareAcronym{APS}{
	short=APS,
	long=azimuth power spectrum
}
\DeclareAcronym{AV}{
	short=AV,
	long=autonomous vehicle
}
\DeclareAcronym{BS}{
	short=BS,
	long=base station
}
\DeclareAcronym{BSM}{
	short=BSM,
	long=basic safety message
}
\DeclareAcronym{CP}{
	short=CP,
	long=cyclic-prefix
}
\DeclareAcronym{DFT}{
	short=DFT,
	long=discrete Fourier transform
}
\DeclareAcronym{DL}{
	short=DL,
	long=downlink
}
\DeclareAcronym{DSRC}{
	short=DSRC,
	long=dedicated short-range communication
}
\DeclareAcronym{FDD}{
	short=FDD,
	long=frequency division duplex
}
\DeclareAcronym{FMCW}{
	short=FMCW,
	long=frequency modulated continuous wave
}
\DeclareAcronym{FoV}{
	short=FoV,
	long=field-of-view
}
\DeclareAcronym{GNSS}{
	short=GNSS,
	long=global navigation satellite system
}
\DeclareAcronym{LIDAR}{
	short=LIDAR,
	long=Light detection and ranging
}
\DeclareAcronym{LOS}{
	short=LOS,
	long=line-of-sight
}
\DeclareAcronym{LPF}{
	short=LPF,
	long=low pass filter
}
\DeclareAcronym{LTE}{
	short=LTE,
	long=long term evolution
}
\DeclareAcronym{MIMO}{
	short=MIMO,
	long=multiple-input multiple-output
}
\DeclareAcronym{MRR}{
	short=MRR,
	long=medium range radar
}
\DeclareAcronym{NLOS}{
	short=NLOS,
	long=non-line-of-sight
}
\DeclareAcronym{NR}{
	short=NR,
	long=new radio
}
\DeclareAcronym{OFDM}{
	short=OFDM,
	long=orthogonal frequency-division multiplexing
}
\DeclareAcronym{ppm}{
	short=ppm,
	long=parts-per-million
}
\DeclareAcronym{RMS}{
	short=RMS,
	long=root-mean-square
}
\DeclareAcronym{RPE}{
	short=RPE,
	long=relative precoding efficiency
}
\DeclareAcronym{RSU}{
	short=RSU,
	long=roadside unit
}
\DeclareAcronym{SNR}{
	short=SNR,
	long=signal-to-noise ratio
}
\DeclareAcronym{UL}{
	short=UL,
	long=uplink
}

\DeclareAcronym{ULA}{
	short=ULA,
	long=uniform linear array
}
\DeclareAcronym{V2I}{
	short=V2I,
	long=vehicle-to-infrastructure
}
\DeclareAcronym{V2V}{
	short=V2V,
	long=vehicle-to-vehicle
}
\DeclareAcronym{V2X}{
	short=V2X,
	long=vehicle-to-everything
}
\DeclareAcronym{VRU}{
	short=VRU,
	long=vulnerable road user
}

\newcommand{\GHz}{\SI{}{\giga\hertz}}

\newcommand{\Ns}{N_{\rms}}

\newcommand{\NRSU}{N_{\rmR\rmS\rmU}}
\newcommand{\NV}{N_{\rmV}}
\newcommand{\MRSU}{M_{\rmR\rmS\rmU}}
\newcommand{\MV}{M_{\rmV}}
\newcommand{\bFBB}{\bF_{\rmB\rmB}}
\newcommand{\bFRF}{\bF_{\rmR\rmF}}
\newcommand{\bWBB}{\bW_{\rmB\rmB}}
\newcommand{\bWRF}{\bW_{\rmR\rmF}}
\newcommand{\alpharc}{\alpha_{r_c}}
\newcommand{\baRSU}{\ba_{\rmR\rmS\rmU}}
\newcommand{\baV}{\ba_{\rmV}}

\newcommand{\bsfRRSU}{\bsfR_{\rmR\rmS\rmU}}
\newcommand{\hbsfRRSU}{\hat{\bsfR}_{\rmR\rmS\rmU}}

\newcommand{\Tr}{T_{\rmr}}
\newcommand{\Tc}{T_{\rmc}}

\DeclareAcronym{mmWave}{short = mmWave, long = millimeter wave}
\DeclareAcronym{MU}{short = MU, long = multi-user}

\begin{document}
%
\title{Deep Learning-based Link Configuration for Radar-aided Multiuser mmWave Vehicle-to-Infrastructure Communication}
\author{Andrew Graff, {\it Student Member, IEEE}, Yun Chen, {\it Student Member, IEEE}, Nuria Gonz\'alez-Prelcic, {\it Senior Member, IEEE}, and Takayuki Shimizu
\thanks{This work was partially supported by a gift from Toyota.}
\thanks{A. Graff is with the Department of Electrical and Computer Engineering, The University of Texas at Austin, Austin, TX 78712, USA \mbox{(e-mail: andrewgraff@utexas.edu)}.}
\thanks{Yun Chen and N. Gonz\'alez-Prelcic are with the Department of Electrical and Computer Engineering, North Carolina State University, Raleigh, NC 27695, USA \mbox{(e-mail: \{ychen273,ngprelcic\}@ncsu.edu)}.}
\thanks{T. Shimizu is with Toyota Motor North America, Mountain View, CA 94043 USA (e-mail: takayuki.shimizu@toyota.com).}
}

\maketitle

\begin{abstract}
Configuring millimeter wave links following a conventional beam training protocol, as the one proposed in the current cellular standard, introduces a large communication overhead, specially relevant in vehicular systems, where the channels are highly dynamic. 
In this paper, we propose the use of a passive radar array to sense automotive radar transmissions coming from multiple vehicles on the road, and a radar processing chain that provides information about a reduced set of candidate beams for the links between the road-infrastructure and each one of the vehicles.  This prior information can be later leveraged by the beam training protocol to significantly reduce overhead.  The radar processing chain estimates both the timing and chirp rates of the radar signals, isolates the individual signals by filtering out interfering radar chirps, and estimates the spatial covariance of each individual radar transmission. Then, a deep network is used to translate features of these radar spatial covariances into features of the communication spatial covariances, by learning the intricate mapping between radar and communication channels, in both line-of-sight and non-line-of-sight settings. The communication rates and outage probabilities of this approach are compared against exhaustive search and pure radar-aided beam training methods (without deep learning-based mapping), and evaluated on multi-user channels simulated by ray tracing. Results show that: (i) the proposed processing chain can reliably isolate the spatial covariances for individual radars, and (ii) the radar-to-communications translation strategy based on deep learning provides a significant  improvement over pure radar-aided methods in both LOS and NLOS channels.
\end{abstract}
\section{Introduction}\label{sec:intro}
\IEEEPARstart{T}{he} automotive industry is experiencing rapid technological advances, producing vehicles that are both more aware of their surrounding and able to communicate with others. This perceptual awareness has been achieved through the use of several onboard sensors, most notably automotive radars. The ability to communicate with infrastructure, referred to as \ac{V2I} communication, allows for the sharing of sensor data, navigation information, multimedia, and more. These applications need high data rates to operate seamlessly.

Wireless communications at \ac{mmWave} bands can achieve such high data rates. Communication at \ac{mmWave} bands, however, typically requires gains through beamforming to reach acceptable signal-to-noise ratios (SNR), often requiring large antenna arrays. While increasing the number of antennas in the array increases the throughput of the system, it comes with a significant hardware cost if each antenna element has its own RF-chain and analog-to-digital converter (ADC). Approaches to reduce the hardware complexity of these large MIMO systems include replacing high-resolution ADCs with low-resolution alternatives, and using a hybrid analog-digital beamforming architecture with fewer RF-chains than the number of antenna elements. While complexity is reduced, these approaches introduce additional challenges. Low-resolution ADCs cause quantization noise, leading to a performance loss. Reducing the number of RF-chains requires hybrid beamforming processing, which can require increased training overhead to form the best digital and analog precoders and combiners for the \ac{mmWave} channel. In addition to the increased training overhead, a hybrid architecture with an analog beamforming stage can only provide a compressed observation of the channel, which further difficults the channel acquisition process for beamforning. In summary, establishing mmWave links with practical MIMO architectures requires a high training overhead. 
This training becomes even more demanding in multiuser \ac{V2I} links, because the channel coherence times are short and users are rapidly entering and exiting the communication cell, resulting in a high probability that users are in initial access. Furthermore, the communication system must be able to maintain a robust link with multiple users at a time, which introduces additional training overhead as well as the potential for interference between users.

The onboard radars on many of these next-generation vehicles provide a unique signal of opportunity for passive radar. With vehicles typically being equipped with multiple automotive radars, often transmitting a predictable \ac{FMCW} waveform also in the mmWave band, a passive radar could extract useful spatial information about the locations of the vehicles. The specific waveform parameters can aid in both, the unique identification of vehicles, as well as the filtering of other interfering radar signals. This ability to filter out interference is especially important, since FMCW automotive signals cause significant interference with each other \cite{Alland2019Interference}. Additionally, and specially relevant for the work presented in this paper, since many automotive radars  operate in a \ac{mmWave} band adjacent to \ac{mmWave} communications bands, the spatial covariance obtained at a passive radar can be similar enough to the spatial covariance of the communication link, and therefore, it can be utilized to reduce the training overhead required to configure the link, as  established in prior work \cite{Ali2020Passive}.

This paper proposes the use of a \ac{RSU} equipped with a passive radar to aid in establishing \ac{MU}-MIMO communication links at \ac{mmWave}. The passive radar senses the transmitted \ac{FMCW} signals from multiple automotive radars at once, isolates the signal from each individual vehicle, estimates the spatial covariance of the individual \ac{FMCW} signals, and then predicts the \ac{mmWave} communication spatial covariance. This predicted communication covariance is then used to extract the main channel directions and select the beams that will act as analog precoders and combiners at the \ac{RSU}, significantly reducing training overhead.

\subsection{Contributions}
The main contributions of this paper are as follows:
\begin{itemize}
\item We propose to leverage the spatial covariance information obtained with a passive radar at the RSU to configure the different millimeter wave communication  links  between the RSU and different connected vehicles that are simultaneously in initial access. 
\item We propose a passive radar processing chain that uses a filter bank architecture to isolate individual \ac{FMCW} signals from a reception containing multiple interfering \ac{FMCW} signals. These isolated signals are then used to estimate the individual spatial covariances corresponding to the radar transmissions coming from different vehicles.
\item We design multiple deep learning architectures to predict the communication link spatial covariance from an estimated noisy radar spatial covariance. These neural networks learn the intricate relations and differences between the spatial covariances in the radar and communication bands. Three variations of neural networks are proposed to predict different functions useful for beamformer design: angular power spectrum (APS) prediction, eigenvector prediction, and covariance vector prediction.
\item We create a ray tracing simulation  of the radar-aided  \ac{MU} vehicular communication scenario and evaluate the performance of the proposed systems in terms of both the sum-rates and outage probabilities of different beam training strategies. This setup is consistent with the 3GPP V2X evaluation methodology for vehicular communication systems, and  emulates commercial vehicles when integrating automotive radar sensing capabilities.
\end{itemize}
\subsection{Prior work}

Out-of-band information to aid mmWave communication \cite{Prelcic2017} can come from several sources, including sub-6 \GHz{} systems \cite{Hashemi2018OOB,Ali2018OOB,Ali2019Covariance}, or sensors such as radar \cite{Ali2020Passive,Prelcic2016Radar,Liu2020Radar}, lidar \cite{Klautau2019Lidar}, inertial-measurement-units \cite{Brambilla2019IMU}, or position information \cite{Kela2016Location,Miao2019UAV}. 

Different approaches that exploit sub-6GHz signals have been proposed to  reduce the beam training time or to estimate the spatial covariance, which is later used to design the beamformers. MmWave link configuration assisted by sub-6 GHz systems has, however, many limitations. 
The use of sub-6 \GHz{} information in \cite{Hashemi2018OOB} is restricted to line-of-sight (LOS) channels. The strategies  in \cite{Ali2018OOB,Ali2019Covariance} is applicable to non-line-of-sight (NLOS) channels, but it requires that both the mmWave and sub-6 \GHz{} channels have identical states (both LOS or both NLOS). Due to the large carrier frequency separation between the two channels, they may have different amounts of obstruction within their Fresnel zones, and thus have different classifications despite propagating through identical environments \cite{Ali2020Passive}. These methods may not work under such circumstances. 

Position information extracted from a Global Positoning System (GPS) has also been used in different ways to reduce overhead of  mmWave link configuration. For example, inverse fingerprinting learns a subset of location-dependent beam-pairs based on past measurements in similar locations, such that with high probability at least one of the vectors in the subset works well \cite{VaEtAlPositionaidedMillimeterWaveV2I2017}\cite{V.VaEtAlInverseMultipathFingerprintingMillimeter2018}\cite{V.VaEtAlOnlineLearningPositionAidedMillimeter2019}\cite{SatyanarayanaEtAlDeepLearningAidedFingerprintbased2019}. Further reductions can happen if there is also knowledge of other connected vehicles (which may have different sizes and act as blockages) \cite{Y.WangEtAlMmWaveVehicularBeamSelection2019} or other context \cite{SimEtAlOnlineContextawareMachineLearning2018a} information.
A channel tracking and beamforming scheme based on position information is proposed in \cite{Kela2016Location}, in which an extended Kalman filter tracks the LOS angles to mobile users to design beamforming weights. Compared to full CSI training, this approach reduced the time-frequency resources required and showed performance improvements which were verified by ray-tracing simulations. Similarly, a beamforming scheme specifically targeted at UAV's using position information is proposed in \cite{Miao2019UAV}, in which the GPS and UAV sensor information is used to estimate positions and direction of arrival (DoA) to improve SINR. Tracking of the complete MIMO channel matrices, also in the context of UAV communication, and leveraging position information to reduce the search space for the channel angular parameters, was also designed in \cite{Javier2018ICC,Javier2019SPAWC}. Beam-tracking for automotive vehicles aided by inertial-measurement-units was proposed in \cite{Brambilla2019IMU}.The common limitation of all these approaches is that they only target LOS scenarios.

A mmWave communication system aided by a lidar in the vehicle is described  in \cite{Klautau2019Lidar}. It was designed to operate only in LOS propagation. Furthermore, lidar requires an active sensor which increases power draw and cost at the roadside unit. In \cite{mashhadi2021federated}, a federated learning scheme was proposed to train a neural network to predict V2I beam selections in both LOS and NLOS scenarios,  using preprocessed spatial information collected from lidar sensors. This work was expanded on in \cite{zecchin2021lidar} by proposing a non-local attention module, a loss function based on knowledge distillation, and a curriculum training approach to target difficult-to-learn scenarios. These advancements improved beam selection accuracy and specifically benefit prediction in NLOS scenarios. These studies demonstrate the capability of neural networks to learn spatial characteristics of communication channels from lidar measurements of the propagation environment.

The first work that proposed leveraging a radar sensor to aid millimeter wave link configuration considers an active radar at the RSU \cite{Prelcic2016Radar} to illuminate receivers on the vehicle and estimate the radar covariance. This is also the first study that experimentally shows there is similarity between the angular information extracted from the radar and the communication
spatial covariances, even when the center frequencies of operation are different.    In \cite{Liu2020Radar}, a dual function radar and communication system was proposed for simultaneously sensing vehicles and establishing the communication link aided by the sensing information. Position information obtained with a radar unit at the road infrastructure was also used in \cite{AliRadarConf2019} to reduce the overhead of the beam training protocol. The accuracy of position information provided by radar is higher than that provided by GPS, what leads to a larger reduction in communication overhead when exploiting position
information provided by a radar sensor than when leveraging GPS-based position, as shown in the field measurements provided in \cite{Graff2019}. Although all these approaches based on an active radar provide an interesting reduction of the link configuration overhead, they only perform well in LOS scenarios. And additional limitation is that the allocation of power to active radar sensing may be prohibitive given a power budget at the roadside unit.  Alternatively, a passive radar approach was taken in \cite{Ali2020Passive}, where the RSU senses signals transmitted from automotive radars onboard the vehicles themselves. This solves the power consumption issue and allows NLOS estimation, but the study was restricted to a single-user case without interference from multiple radars. Furthermore, there is an inherent mismatch between the estimated radar covariance and the true communication covariance due to different operation frequencies or different locations of the radars and communication transceivers in the vehicles. In this paper, we overcome these limitations by building upon our preliminary work in \cite{ChenGlobecom2021} to add multiuser capabilities and to further refine the covariance estimate by translating the radar covariance to the communication domain. To this aim we use neural networks which effectively learn the mismatches between radar and communication channels. Although the work in \cite{ChenGlobecom2021} already explores the  idea of learning mismatches, only a single user scenario and the estimation of the APS are considered. 


\textbf{Notation:} We use the following notation throughout the paper. Bold lowercase $\bx$ is used for column vectors, bold uppercase $\bX$ is used for matrices, non-bold letters $x$, $X$ are used for scalars. $[\bx]_i$ and $[\bX]_{i,j}$, denote $i$th entry of $\bx$ and entry at the $i$th row and $j$th column of $\bX$, respectively. We use serif font, e.g., $\bsfx$, for the frequency-domain variables. Superscript $\transp$, $\ast$ and $\dagger$ represent the transpose, conjugate transpose and pseudo inverse, respectively. $\bzero$ and $\bI$ denote the zero vector and identity matrix respectively. $\cC\cN(\bx,\bX)$ denotes a complex circularly symmetric Gaussian random vector with mean $\bx$ and covariance $\bX$, and $\cU[a,b]$ is a Uniform random variable with support $[a,b]$. We use $\bbE[\cdot]$ and $\|\!\cdot\!\|_\rmF$ to denote expectation and Frobenius norm, respectively.

\section{System model}\label{sec:sig_model}

\begin{figure}[h!]
\centering
\includegraphics[width=\columnwidth]{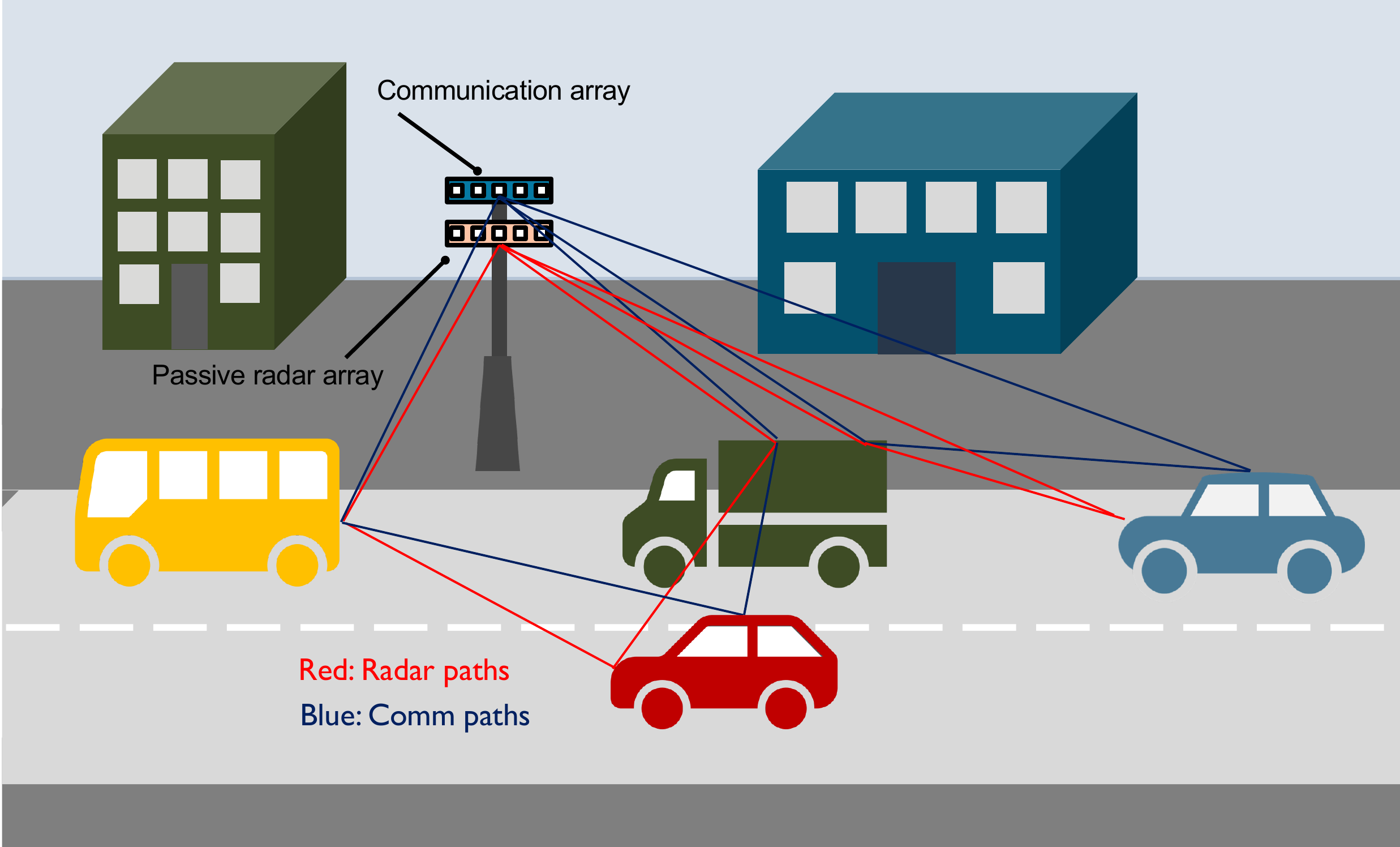}
\caption{A multiuser V2I communication system where the RSU is equipped with a passive radar array to tap the automotive radar signals coming from the connected vehicles.}
\label{fig:system_model_MU}
\end{figure}

We consider the MU-MIMO V2I communication system represented in Fig.~\ref{fig:system_model_MU}, where the \ac{RSU} is located on the side of a roadway and several ego-vehicles are driving along the road with other non connected vechicles. The \ac{RSU} is equipped with a passive radar \ac{ULA} and a communications \ac{ULA}. The ego vehicles have 4 \ac{ULA}s for communications and 4 single-antenna automotive radars. The communication arrays in the vehicles are placed in accordance with \ac{3GPP} proposals~\cite{3GPP37885}, and the radar arrays are placed at the 4 corners of the vehicle as in many commercial models. The passive radar array at the \ac{RSU} will use receptions of the automotive radar signals to estimate the radar spatial covariances for each link. This covariances will then be used to configure the MU-MIMO mmWave communication link.

\subsection{Communication system model}\label{sec:comm}

The communication array on the \ac{RSU} is equipped with $\NRSU$ antennas and $\MRSU\leq\NRSU$ RF-chains. We let $A$ denote the number of communication arrays at the ego vehicle. Each vehicle array has $\NV$ antenna elements and $\MV\leq\NV$ RF-chains. This hybrid architecture supports $\Ns\leq\min\{\MRSU,\MV\}$ data-streams. The communication link is based on a $K$ sub-carrier \ac{OFDM} system, with modulated symbols $\bsfs[k]\in\bbC^{\Ns\times1}$ such that $\bbE[\bsfs[k]\bsfs^\ast[k]]=\frac{P_\rmc}{K\Ns}\bI_{\Ns}$ and  $P_\rmc$ denotes the total average transmitted power. The baseband precoder $\bFBB[k]\in\bbC^{\MRSU\times\Ns}$ and RF precoder $\bFRF\in\bbC^{\NRSU\times\MRSU}$ are combined to form the hybrid precoder $\bF[k]=\bFRF\bFBB[k]\in\bbC^{\NRSU\times\Ns}$ on sub-carrier $k$. The RF precoder is realized using quantized phase shifters and is the same across all subcarriers. Letting $\zeta_{i,j}$, $i=1,\ldots,\NRSU$, $j=1,\ldots,\MRSU$, be the quantized phase shift, the RF precoder is described as $[\bFRF]_{i,j}=\frac{1}{\sqrt{\NRSU}}e^{\compj \zeta_{i,j}}$. The total power constraint is enforced as $\sum_{k=1}^{K}\|\bF[k]\|_\rmF^2=K\Ns$. 

The baseband combiner $\bWBB^{(a)}[k]\in\bbC^{\MV\times\Ns}$ and RF combiner $\bWRF^{(a)}\in\bbC^{\NV\times\MV}$ are multiplied to form the hybrid combiner $\bW^{(a)}[k]=\bWRF^{(a)}\bWBB^{(a)}[k]\in\bbC^{\NV\times\Ns}$ on sub-carrier $k$. The $\NV\times\NRSU$ frequency-domain MIMO channel at array $a\in{A}$ is denoted as $\bsfH^{(a)}[k]$. Assuming perfect synchronization, the received signal on sub-carrier $k$ after processing is
\begin{equation}
\bsfy^{(a)}[k]=\bsfW^{(a)\ast}[k]\bsfH^{(a)}[k]\bsfF[k]\bsfs[k]+\bsfW^{(a)\ast}[k]\bsfn^{(a)}[k],
\label{eq:rxpost}
\end{equation}
where $\bsfn^{(a)}\sim\cC\cN(\bzero,\sigma_{\bsfn}^2\bI)$ is additive white Gaussian noise.

\subsection{Channel model} 

The wideband channel is modeled geometrically with $C$ clusters. Each of the clusters experiences a mean time delay $\tau_c \in \bbR$, mean \ac{AoA} $\theta_c \in [0,2\pi)$, and mean \ac{AoD} $\phi_c \in [0,2\pi)$. Assuming there are $R_c$ paths in each cluster, each path $r_c\in[R_c]$ has complex gain $\alpha_{r_c}$, relative time-delay $\tau_{r_c}$, relative arrival angle shift $\vartheta_{r_c}$, and relative departure angle shift $\varphi_{r_c}$. The array response vectors are $\baRSU(\phi)$ at the \ac{RSU} and $\baV(\theta)$ at the ego-vehicle. The uniform spacing between array elements is $\Delta$, normalized to units of wavelength. The \ac{RSU} response vector and ego-vehicle response vectors are defined as

\begin{equation}
	\baRSU(\theta)=[1,e^{\compj 2\pi\Delta\sin(\theta)},\cdots,e^{\compj(\NRSU-1) 2\pi\Delta\sin(\theta)}]^\transp.
\end{equation}

\begin{equation}
	\baV(\phi)=[1,e^{\compj 2\pi\Delta\sin(\phi)},\cdots,e^{\compj(\NV-1) 2\pi\Delta\sin(\phi)}]^\transp.
\end{equation}

We will remove the notation $(a)$ in the channel $\bH$ for the following equations. We will define the analog filtering and pulse shaping effect at delay $\tau$ as $p(\tau)$. $\Tc$ will denote the signaling interval. The delay-$d$ \ac{MIMO} channel matrix $\bH[d]$ is~\cite{Alkhateeb2016Frequency}

\begin{align}
	\bH[d]=\sum_{c=1}^C \sum_{r_c=1}^{R_c} \alpharc p& (d\Tc-\tau_c-\tau_{r_c})\times\nonumber\\
	&\baV(\phi_c+\varphi_{r_c})\baRSU^\ast(\theta_c+\vartheta_{r_c}).
	\label{eq:timedomch}
\end{align}

If there are $D$ delay-taps in the channel, the channel at sub-carrier $k$, $\bsfH[k]$ is~\cite{Alkhateeb2016Frequency}
\begin{equation}
	\bsfH[k]=\sum_{d=0}^{D-1}\bH[d] e^{-\compj \tfrac{2\pi k}{K}d}.
	\label{eq:freqdomch}
\end{equation}

\subsection{Covariance model}\label{sec:covmod}

We define the spatial covariance at the \ac{RSU} on sub-carrier $k$ as $\bsfRRSU[k]=\frac{1}{\NV}\bbE[\bsfH^\ast[k]\bsfH[k]]$. By assuming that the covariance does not change across sub-carriers~\cite{Bjornson2009Exploiting}, we can create an estimate by averaging over all sub-carriers $\hbsfRRSU=\frac{1}{K}\sum_{k=1}^K \hbsfRRSU [k]$. Since we will only use covariance estimates to design the analog precoder and combiner, this is an appropriate assumption, as the baseband precoder and combiner will be designed independently and will account for sub-carrier-dependent covariance variations~\cite{Ali2020Passive}.

\subsection{Radar system model}\label{sec:rad}

Each automotive radar in the environment transmits a unique \ac{FMCW} signal. We will assume there are $M$ radars transmitting. The $m$th radar, for $m\in[M]$, has a chirp rate of $\beta_m$, a time offset of $\Delta t_m$, and a phase offset of $\Delta \phi_m$. We will assume all radars operate with the same bandwidth $B$. Then the chirp period is defined as $T_m = \frac{B}{\beta_m}$.

Then the transmitted signal can be defined as
\begin{multline}
s_m(t) = \sqrt{P_\rmr}\exp{\left(j2\pi\left(f_rt+\frac{\beta{}_m t^2}{2}\right) + j\phi_m \right)} \\
 \text{for } t\in[\Delta t_m, \Delta t_m + T_m].
\end{multline}
This transmitted signal repeats every $T_m$ seconds. The received signal on the $N_\rmr$ element antenna array on the \ac{RSU} will be denoted as a vector $\bx(t)\in \bbC^{N_\rmr}$. Assume that due to multipath effects, the radar transmission propagates along $R_m$ paths. Each path $r_m \in [R_m]$ experiences an attenuation of $\alpha_{r_m}$ and a time delay of $\tau_{n,r_m}$ during propagation to the $n$th antenna. The received signal at antenna $n$ is
\begin{equation}
[\bx(t)]_n = \sum_{m=1}^{M} \sum_{r_m=1}^{R_m} \alpha_{r_m} s(t-\tau_{n,r_m}).
\end{equation}

We can model the propagation delay as the sum of two components: one accounting for common distance $\tau$ and another accounting for the difference among antenna elements at the \ac{ULA} $\tau^\prime_n$. This delay at antenna $n$ is described as $\tau_{n,r_m}=\tau_{r_m} + \tau^\prime_{n,r_m}$~\cite{Katkovnik2002High}. We assume our \ac{ULA} has half-wavelength spacing and that the signal from radar $m$ and path $r_m$ arrives at an angle of $\theta_{r_m}$, so

\begin{equation}
\tau^\prime_{n,r_m}=\frac{\sin \theta_{r_m} (n-1) }{2 f_\rmr}.
\end{equation}

Then we collect the $I$ samples of the signal into a matrix $\bY\in\bbC^{N_\rmr \times I}$. Let $i\in\{1,2,\cdots,I\}$ denote the sample index, and $\Tr$ denote the sampling time. Then the $i$th sample on the $n$th antenna is $[\bY]_{n,i}=[\bx(i \Tr)]_n$. The spatial covariance of the received radar signal could then be estimated as $\hat{\bm{R}}=\frac{1}{I}\bY\bY^\ast$.

However, this covariance estimate is not particularly useful when multiple vehicular radars are transmitting. The covariance will contain all of the interfering signals from all $M$ vehicular radars that are transmitting. As a result, the \ac{APS} computed from such an estimate may be dominated by the contributions from higher SNR signals while the contributions from low SNR signals are undetectable due to the interference and sidelobes. A more useful covariance would be the isolated covariance of each transmitting radar signal. Let us define the true spatial covariance for signal $m$. This will be done by propagating a unit power dirac-delta signal through the radar channel from radar $m$, i.e.

\begin{equation}
    [\check{\bx}_{m}(t)]_n = \sum_{r_m=1}^{R_m} \alpha_{r_m} \delta(t-\tau_{n,r_m}).
\end{equation}

\noindent
Then perform the same sampling and covariance estimation as before:

\begin{equation}
[\check{\bY}_m]_{n,i}=[\check{\bx}_{m}(i \Tr)]_n,
\end{equation}

\begin{equation}
\bm{R}_m=\frac{1}{I}\check{\bY}_m\check{\bY}_m^\ast.
\end{equation}

\noindent
$\bm{R}_m$ is our ideal isolated spatial covariance from the radar signal $m$. The next section will explore the signal processing chain designed to creates estimates $\hat{\bm{R}}_m$ of the isolated spatial covariance from the total received signal $\bx(t)$.

\subsection{Multiuser Separation}\label{sec:mu_sep}
As mentioned, the received signal $\bY$ contains contributions from all transmitting radars in the environment. However, the \ac{RSU} must now attempt to estimate the individual radar covariances for each vehicle, which will later be used to assist in precoder design for downlink communications. We assume that for vehicles in initial access, the \ac{RSU} has no knowledge of the automotive radar chirp rate or timing. Furthermore, the \ac{RSU} may have no knowledge whether or not a new vehicle has entered its area-of-coverage and must be able to detect radar transmissions from new vehicles. As such, we propose a FMCW mixing filter bank processing chain to detect vehicles, suppress interference from other automotive radar signals, and estimate the spatial covariance for each vehicle in the area-of-coverage individually.

\begin{figure}[h!]
\centering
\includegraphics[width=\columnwidth]{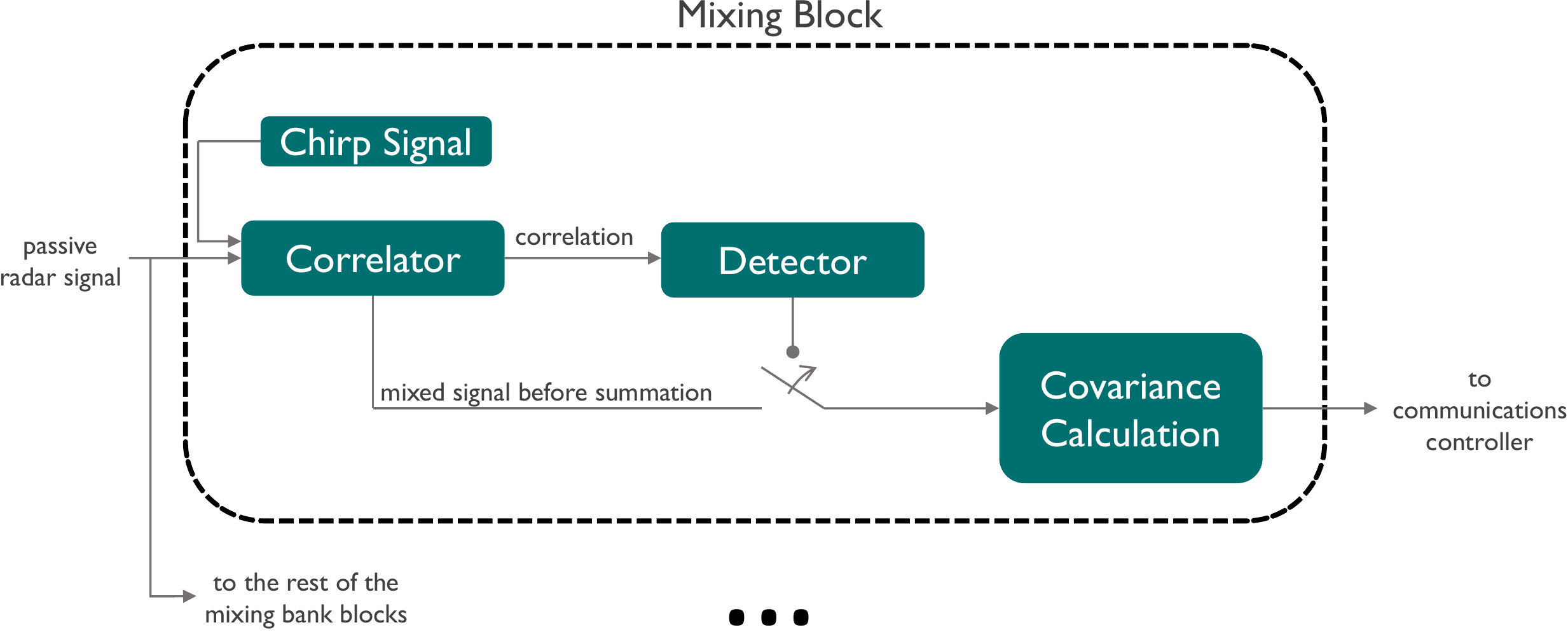}
\caption{Visualization of a single mixing block within the mixing bank. The received signal is mixed with a chirp signal, sampled, and correlated. The detector then determines whether a FMCW signal is present corresponding to the block's chirp rate and finds the chirp timing. The covariance is then estimated based on the mixed signal. }
\label{fig:mixing_block}
\end{figure}

\begin{figure*}[t!]
\centering
\includegraphics[width=0.8\textwidth]{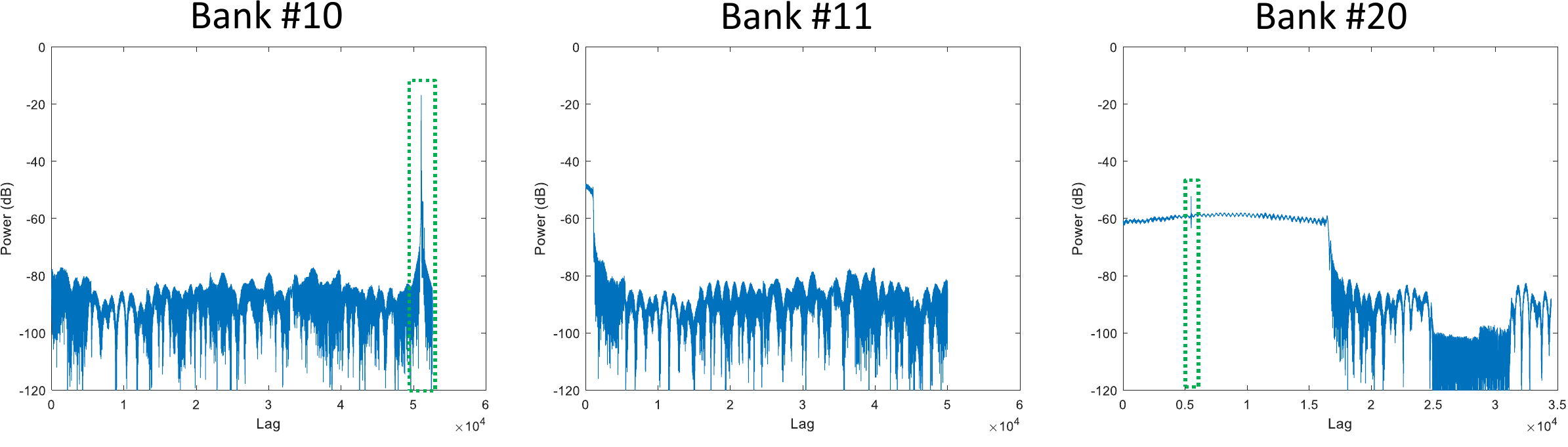}
\caption{The correlator output of 3 mixing blocks with reference FMCW chirp rates of 11, 12, and 19 MHz/$\mu$s. Sharp peaks are shown in the first and last output corresponding to passive radar receptions of signals with a matching chirp rate. The second output has no such peak.}
\label{fig:mixing_output}
\end{figure*}

For now, let's consider a single block within the mixing bank, which is visualized in Fig.~\ref{fig:mixing_block}. The received signal is mixed with a reference FMCW signal $s_{\text{ref}}(t)$ with the desired chirp rate $\beta_\text{mix}$. Much like the transmitted FMCW signals, the reference FMCW signal has a chirp period of $T_\text{mix}$ and repeats every $T_\text{mix}$ seconds, 
\begin{equation}
    s_{\text{ref}}(t) = \exp{\left(-j2\pi\left(f_rt+\frac{\beta_r t^2}{2}\right)\right)} \quad \text{for } t\in[0, T_m].
\end{equation}
The output mixed signal is denoted as $\mathbf{x_{\text{mix}}}(t)$, and can be written as
\begin{equation}
    \mathbf{x_{\text{mix}}}(t) = \bx(t)s_{\text{ref}}(t).
\end{equation}
This signal after sampling is then called $\mathbf{Y}_\text{mix}$, and can be expressed as
\begin{equation}
    [\mathbf{Y}_\text{mix}]_{n,i} = [\mathbf{x_{\text{mix}}}(iT_r)]_{n,i}.
\end{equation}
Note that the received chirp signals are not time-aligned with the reference chirp signal. The output of this mixer is then sampled and processed digitally.

The next stage in the processing block is the correlator. The sampled signal is then digitally mixed with an offset-correction signal before the $I$ samples are summed together. This digital mixing and summation is repeated for every lag. The output of the correlator is then passed to a detector. The digital correction signal for lag $l$ is defined as
\begin{equation}
    [S_{\text{corr}}]_{l,i} = exp{\left(j2\pi\left(\frac{\beta_m (lT_r)^2}{2}iT_r\right)\right)}.
\end{equation}
The corrected sampled mixed signal at lag $l$ is defined as
\begin{equation}
    [\mathbf{Y}_{\text{mix},l}]_{n,i} = ([\mathbf{Y}_\text{mix}]_{n,i})([S_{\text{corr}}]_{l,i}).
\end{equation}
And finally, the correlator output is the defined as
\begin{equation}
    [\mathbf{C}]_{n,l} = \sum_{i=1}^{I} [\mathbf{Y}_{\text{mix},l}]_{n,i}.
\end{equation}
The digital correction signal accounts for the frequency of the reference FMCW signal at the start of lag $l$. This makes the correlator output equivalent to mixing and summing with an FMCW signal that starts at frequency $f_r$ at every lag $l$. This saves on hardware complexity by allowing each mixing block to only require mixing with a single FMCW reference signal. The digital correction is then an element-wise complex multiplication. Assuming that the FMCW reference signal is controlled by a voltage-controlled oscillator (VCO), the digital processor can have knowledge of the reference signals frequency at each sample time.

When the chirp start and chirp rate of the block align with the chirp start and chirp rate of a received signal, the magnitude of the output of the correlator will exhibit a sharp peak. For chirp rates that do not align, the output of the correlator will have its power spread out over the lag domain. To show this, consider a case where 51 mixing blocks are used in the filter bank, each with a uniformly spaced chirp rates between 10 and 60 MHz/$\mu$s. This is visualized in Fig.~\ref{fig:mixing_output}, where the correlator output of the same received signals is shown for 3 of the 51 mixing blocks. In bank \#10 and bank \#20, sharp peaks exist because the received signal contains FMCW pulses matched to their mixing chirp rates of 11 and 19 MHz/$\mu$s. Bank \#11 has no such peak, because the received signal contained no FMCW pulses with a chirp rate of 12 MHz/$\mu$s. In bank \#11, the power of the chirp detected in bank \#10 becomes spread out over the lag domain. This spreading effects grows as the signal and mixing chirp rates become further separated. In bank \#20, the chirp detected in bank \#10 is spread out significantly and can be seen as the plateau at approximately -60 dB, allowing for the much lower power signal at 19 MHz/$\mu$s to be detected. The remaining interference from the spread out signal is not present across the entire lag domain however, so the detector of these peaks must be able to adapt to changing interference and noise levels. The objective of the detector is to determine whether such a peak exists in the output of the correlator and to remain robust against the interference power from other signals that have different chirp starts and chirp rates.

\begin{figure}[h!]
\centering
\includegraphics[width=\columnwidth]{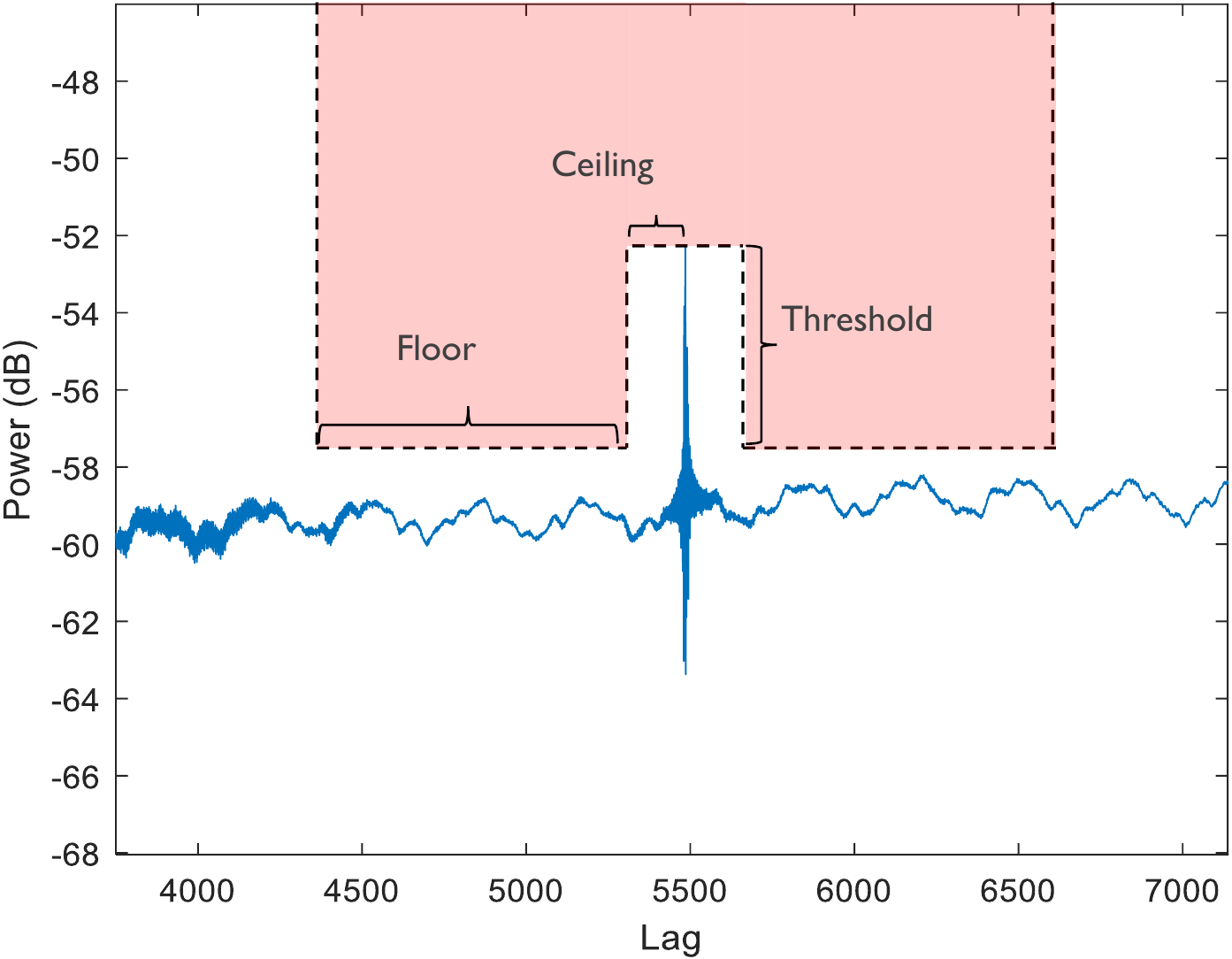}
\caption{A visualization of the max CFAR detector. For each lag, the boundary box defined by the guard cells and floor cells is defined relative to the power at the lag being tested. If the signal does not conflict with the boundary box, a detection is marked at that particular lag.}
\label{fig:detecor}
\end{figure}

For our purposes, we propose the use of max CFAR (constant false alarm rate) detector. The interference power at any given time delay and mixing block may not be known beforehand, especially as vehicle enter and exit the area-of-coverage. As such, an adaptive detection algorithm that estimates the interference power and determines a threshold dynamically is desirable. The max CFAR detector is visualized in Fig.~\ref{fig:detecor}. Max CFAR estimates the noise and interference power around a cell-under-test (CUT) by taking the maximum power of a set of cells that neighbor the CUT. A set of guard cells close to the CUT are ignored in this estimation, because power from the CUT may leak into close by cells. However, we enforce that the power at the CUT is greater than all the powers in the guard cells. In our application, the set of guard cells must account for the delay spread of the radar channel. Once the noise and interference power has been estimated, a detection threshold is determined by multiplying the power by some scaling factor. This scaling factor can be tuned by the system designer to achieve a desired false alarm rate. If the power in the CUT exceeds this threshold, the system considers this a detection. Max CFAR can be implemented efficiently digitally with an FPGA (see \cite{Cumplido2004} for example).

Let the $N_{\text{guard}}$ be the number of guard cells, $N_{\text{floor}}$ be the number of cells beyond the guard cells that are used to estimate the noise and interference power, and $P_{\text{det}}$ be the power threshold for detection. Define the set of guard cell offsets as $\Delta L_{\text{guard}} = [-N_{\text{guard}},1] \bigcup [1,N_{\text{guard}}]$. Define the set of floor cell offsets as $\Delta L_{\text{floor}} = [-N_{\text{guard}}-N_{\text{floor}},-N_{\text{guard}}] \bigcup [N_{\text{guard}},N_{\text{guard}}+N_{\text{floor}}]$. Assume that the correlator output at lag $l$ is our CUT. We decide to take the maximum power across antennas in these detections steps as well. The CUT power $P_{\text{CUT}}$ is defined as
\begin{equation}
    P_{\text{CUT}} = \max_n |[\mathbf{C}]_{n,l}|^2.
\end{equation}
The noise and interference estimate is defined as
\begin{equation}
    P_{\text{floor}} = \max_n \max_{\Delta l \in \Delta L_{\text{floor}}} |[\mathbf{C}]_{n,l+\Delta l}|^2.
\end{equation}
The power in the guard cells is also defined as
\begin{equation}
    P_{\text{guard}} = \max_n \max_{\Delta l \in \Delta L_{\text{guard}}} |[\mathbf{C}]_{n,l+\Delta l}|^2.
\end{equation}
The CUT is then marked as a detection if the following conditions are met:
\begin{equation}
    \begin{split}
        P_{\text{CUT}} > P_{\text{guard}}, \\
        P_{\text{CUT}} > P_{\text{det}}P_{\text{floor}}.
    \end{split}
\end{equation}

Let $D$ be the set of lags where detections are found. The sampled output of the mixer corresponding to these detected lags, $[\mathbf{Y}_{\text{mix},l}]_{n,i} \forall l \in D$, is passed to the covariance estimator. It should be clarified that the samples passed to the covariance estimator are samples before the summation operation in the correlator. By detecting the correct time delay and chirp rate of a particular FMCW reception and then mixing with a reference chirp corresponding to these exact parameters, we experience a power gain in this particular FMCW signal and a suppression of the other interfering FMCW signals. Furthermore, this power is concentrated in spikes near DC since the correct lag and time delay have been estimated. Therefore, a lowpass filter with a bandwidth $B_f$ greater than the multipath spread of the propagation channel can be applied to the detected signals to filter out the interfering radar transmissions.

\begin{equation}
    \hat{\bY}_l = \text{lowpass}\{\mathbf{Y}_{\text{mix},l}\}
\end{equation}

Assuming the detection is accurate, this process isolates the received FMCW signals from each vehicle, allowing the \ac{RSU} to estimate each vehicle's radar spatial covariance separately. These spatial covariances are then subsequently passed to the communications controller, which will establish a link with the newly detected vehicles in the area-of-coverage.

\section{Covariance Prediction}\label{sec:cov_pred}

As per the previous section, the detector outputs a set of sampled post-mixing signals $[\mathbf{Y}_{\text{mix},l}]_{n,i} \forall l \in D$ that are filtered out to obtain $\hat{\bY}_l$. The radar spatial covariance can be estimated independently for each of these signals
\begin{equation}
\hat{\bm{R}}_{l}=\frac{1}{I}\hat{\bY}_l\hat{\bY}_l^\ast.
\end{equation}
The filtering and mixing results in a large gain in the SIR, but some interference power may still remain. Assuming that detection $l$ corresponds to the vehicle our system is establishing a communication link with, The controller must now estimate $\bsfRRSU$ from the noisy radar covariance estimate $\hat{\bm{R}}_{l}$. A neural network will be trained and applied to handle this mapping between the radar and communication domains. However, direct prediction of the complete covariance matrices ignores much of the spatial structure and requires unreasonably high dimensional inputs and outputs. Instead, we train neural networks to input and predict three lower-dimensional features related to the covariance: the APS, the dominant eigenvector, and the covariance vector, which is obtained as the first column of the covariance matrix.

\begin{figure*}[t!]
\centering
\includegraphics[width=0.9\textwidth]{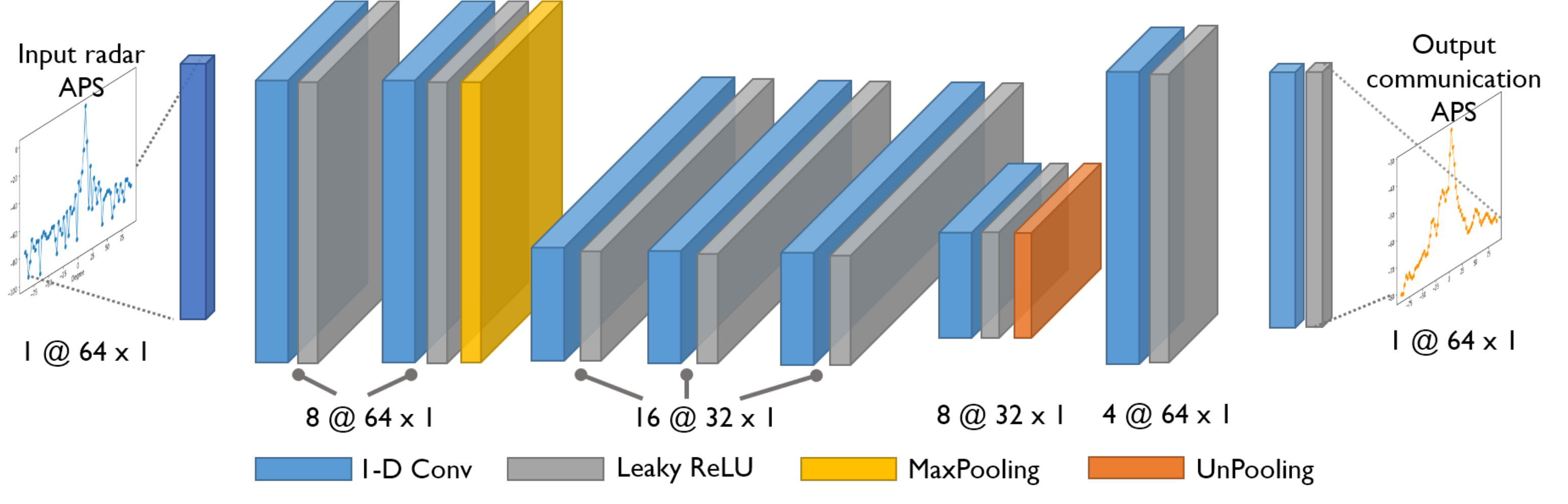}
\caption{The neural network architecture for APS predictions.}
\label{fig:APS_pred_NN_arch}
\end{figure*}

A straightforward way to realize radar-to-communication mapping is to extract the APS from the radar and the communication covariance matrices using the DFT, and use them to train a network that predicts the communication APS using the input radar APS as described in the preliminary work \cite{ChenGlobecom2021}. Once the network has been trained, only the radar APS is needed to predict the communication APS, that will be used to design the beamformers. As the APS is the power distribution over all the directions of arrival, the shape of the APS contains observable features, e.g, peaks and saddles. If the DFT beamforming matrix is not oversampled, this reduces the dimensionality of the prediction from $N^2$ to $N$.

An alternative approach consists os using the dominant eigenvector of the estimated radar covariance to predict the dominant eigenvector of the communication covariance. This serves two purposes. First, it reduces the dimensionality of the prediction from $N^2$ to $N$, making the training and implementation of the neural network significantly simpler. Second, it can help further isolate the signal of interest from the remaining interference after isolation and filtering. In our communication protocol, which will be described in detail in Section \ref{sec:simsetup}, only a single stream will be transmitted to each target to ensure each link has the highest possible SNR. Therefore, predicting the dominant eigenvector of the communication covariance will naturally approximate the spatial weights corresponding to this stream. Consider the eigendecompositions $\hat{\bm{R}}_{l} = \bQ_l\mathbf{\Lambda}_l\bQ_l^{-1}$ and $\bsfRRSU = \bQ_{\text{RSU}}\mathbf{\Lambda}_{\text{RSU}}\bQ_{\text{RSU}}^{-1}$, where the columns of $\bQ_l$ and $\bQ_{\text{RSU}}$ are the eigenvectors of each covariance, and $\mathbf{\Lambda}_l$ and $\mathbf{\Lambda}_{\text{RSU}}$ are diagonal matrices containing the eigenvalues of each covariance. Let $\bv_l$ be the eigenvector in $\bQ_l$ corresponding to the greatest eigenvalue in $\mathbf{\Lambda}_l$. Let $\bv_\text{RSU}$ be the eigenvector in $\bQ_\text{RSU}$ corresponding to the greatest eigenvalue in $\mathbf{\Lambda}_\text{RSU}$. The neural network $\mathcal{N}_{\text{eig}}(\cdot)$ will take $\bv_l$ as input and predict $\hat{\bv}_\text{RSU}$.

Finally, another alternative relies on translating the radar covariance vector to the communication covariance vector \cite{ChenGlobecom2021}. In this case, the special structure of the covariance matrix is leveraged to reduce dimensionality. Toeplitz completion cite{Ali2020Passive, higham2002computing} is first used to project the measured covariance matrix to the Toeplitz, Hermitian and positive semi-definite cone $\bT_+^{N}$, i.e.,  
\begin{equation}
    \widetilde{\bR}_l(\widetilde{\bR}_{\text{RSU}})=\arg\min\limits_{\bX\in\bT_+^N}||\bX-[\hat{\bR}_l(\widetilde{\bR}_{\text{RSU}})-\sigma_n^2\bI]||_\text{F},
\end{equation}
where $\widetilde{\bR}_l$ and $\widetilde{\bR}_{\text{RSU}}$ are the projected covariance matrix for the radar and communication channels, which could be fully represented by their first columns, denoted as $\widetilde{\br}_l, \widetilde{\br}_{\text{RSU}}\in\mathbb{C}^{N_\text{RSU}\times 1}$ and called covariance vectors. This projection keeps most of the information of a covariance matrix as it approximates the closest Toeplitz matrix during iterations. Then, a neural network $\mathcal{N}_{\text{col}}(\cdot)$ is adopted for covariance vector predictions to realize radar-to-communication channel translations. This method also reduces the prediction dimension from $N^2$ to $N$.

\subsection{Neural network architectures}

Different network architectures are designed in this section to map the different radar features to features in the communication domain.
Fig. \ref{fig:APS_pred_NN_arch} shows the network designed to predict the communication APS from the radar APS.
1D convolutional layers are well-known for extracting local features and are adopted in the proposed NN. LeakyReLU is used as the  activation  function  in  the  network to help avoid both dying ReLU and vanishing gradient problems. The loss function defines how error in the predictions is penalized and how the gradients are determined when optimizing the network. The MSE between the predicted and true communication APS is treated as the loss function during the training.

\begin{figure}[h!]
\centering
\includegraphics[width=\columnwidth]{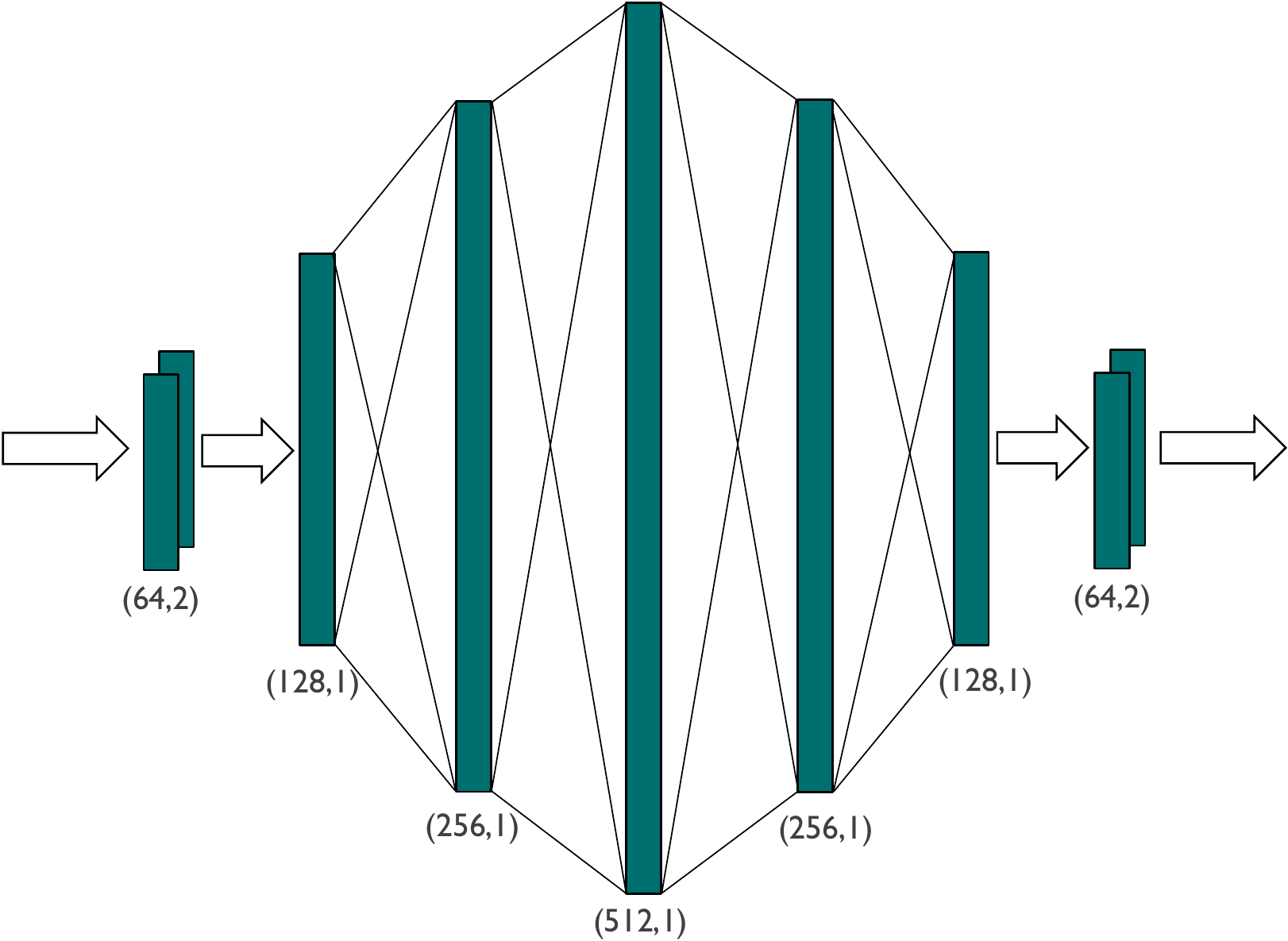}
\caption{The neural network architecture that is trained to predict communication link spatial eigenvectors from radar spatial eigenvector estimates.}
\label{fig:neural_network_eigen}
\end{figure}

The eigenvector prediction neural network has to handle complex data. The input to the network is a 64 element vector containing the complex values of the input eigenvector, as illustrated in Fig.~\ref{fig:neural_network_eigen}. This complex input vector is then converted from complex values to magnitude and phase components. The magnitude and phase components are then stacked together in a single vector of 128 elements. After this restructuring, the vectorized data passes through 5 fully-connected layers containing 128, 256, 512, 256, and 128 activation units each. Each activation unit uses a LeakyReLU activation function with a shape parameter of $\alpha=0.1$. A dropout layer is also placed after the activations of the 3rd fully-connected layer, with a dropout rate of 50\%. The output of the last layer is then scaled to unit norm before being output. The output data is a 128 element real-valued vector where the first 64 elements correspond to the real components of the predicted eigenvector and the last 64 elements correspond to the imaginary components of the predicted eigenvector. For optimizing the network, the loss function is chosen again as the MSE between the predicted and true communication APS, so the beamformer based on this eigenvector points to the main channel angular direction. 

The final network design is suitable for mapping the radar covariance vector to the communication covariance vector. $\widetilde{\br}_l$ and $\widetilde{\br}_{\text{RSU}}$ contain both real and imaginary parts, so the network should also take a two-channel input and output the predicted real and imaginary part of $\widetilde{\br}_{\text{RSU}}$, i.e.,
\begin{equation}
    [\mathfrak{R}\{\hat{\br}_{\text{RSU}}\}, \mathfrak{I}\{\hat{\br}_{\text{RSU}}\}]= \mathcal{N}_{\text{col}}([\mathfrak{R}\{\widetilde{\br}_l\}, \mathfrak{I}\{\widetilde{\br}_l\}];\bu),
\end{equation}
where $\bu$ is the network parameter to be trained. As the covariance column is structure agnostic, and the real and imaginary parts are processed separately, special assumptions (e.g., the input follows any distributions or there are observable spatial features) are not necessary for the input. As such, layers that serve specific functions like extracting local features or keeping historical memories are not used. Fully-connected layers, which allow learning from all the combinations of the features embedded in the covariance, are still suited for this situation. Considering the complexity and effectiveness of the prediction, the network architecture used for eigenvector prediction can be reused with slight modifications. The middle layer containing 512 neurons is neglected, and dropout is not used as the network is not very deep. Here we use Tanh \cite{kalman1992tanh} to constrain the passing values to be in $[-1,1]$. A new cost function is proposed for the purpose of learning spatial eigenvectors. This loss function is shown in Algorithm \ref{alg:aps_loss}. The loss function computes the mean squared error of between the predicted eigenvector's APS and the true eigenvector's APS. Each APS is computed using a Fast-Fourier-Transform (FFT) with a 35 dB Chebyshev windowing function applied. The predicted covarinace vector is transformed back to the Toeplitz covariance matrix $\widetilde{\bR}(\hat{\br}_\text{RSU})$, then the loss could be calculated by Equ.~(\ref{loss_col}):
\begin{equation}\label{loss_col}
	\mathcal{L}({\rm \bu})=\mathbb{E}\{||diag(\bF^*\widetilde{\bR}(\hat{\br}_\text{RSU})\bF)-\bd_c\},
\end{equation}
where $\bd_c$ is the true communication APS.

\begin{algorithm}
	\caption{APS\_loss}
	\begin{algorithmic}
		\STATE Inputs:
		\STATE $\bm{v}_{\text{pred}} \in\mathbb{C}^{N\times{}1}$ - Predicted eigenvector
		\STATE $\bm{v}_{\text{true}} \in\mathbb{C}^{N\times{}1}$ - True eigenvector
		\STATE
		\STATE Begin:
		\STATE $\bm{c} \leftarrow \text{chebwin(N,35)}$
		\STATE $\bm{z}_\text{pred} \leftarrow |\text{FFT}(c\odot\bm{v}_{\text{pred}})|^2$
		\STATE $\bm{z}_\text{true} \leftarrow |\text{FFT}(c\odot\bm{v}_{\text{true}})|^2$
		\STATE $\text{loss} \leftarrow \frac{1}{N} \sum_{n=1}^{N} |[\bm{z}_\text{pred}]_n - [\bm{z}_\text{true}]_n|^2$
		\RETURN $\text{loss}$
		
	\end{algorithmic}
	\label{alg:aps_loss}
\end{algorithm}

\section{Simulation results}\label{sec:simres} 
We will now present simulation data demonstrating the utility of multiuser covariance separation, machine-learning based covariance prediction, and our downlink MU-MIMO communication scheme that uses these new methods to reduce training overhead.

\begin{figure}[h!]
\centering
\includegraphics[width=\columnwidth]{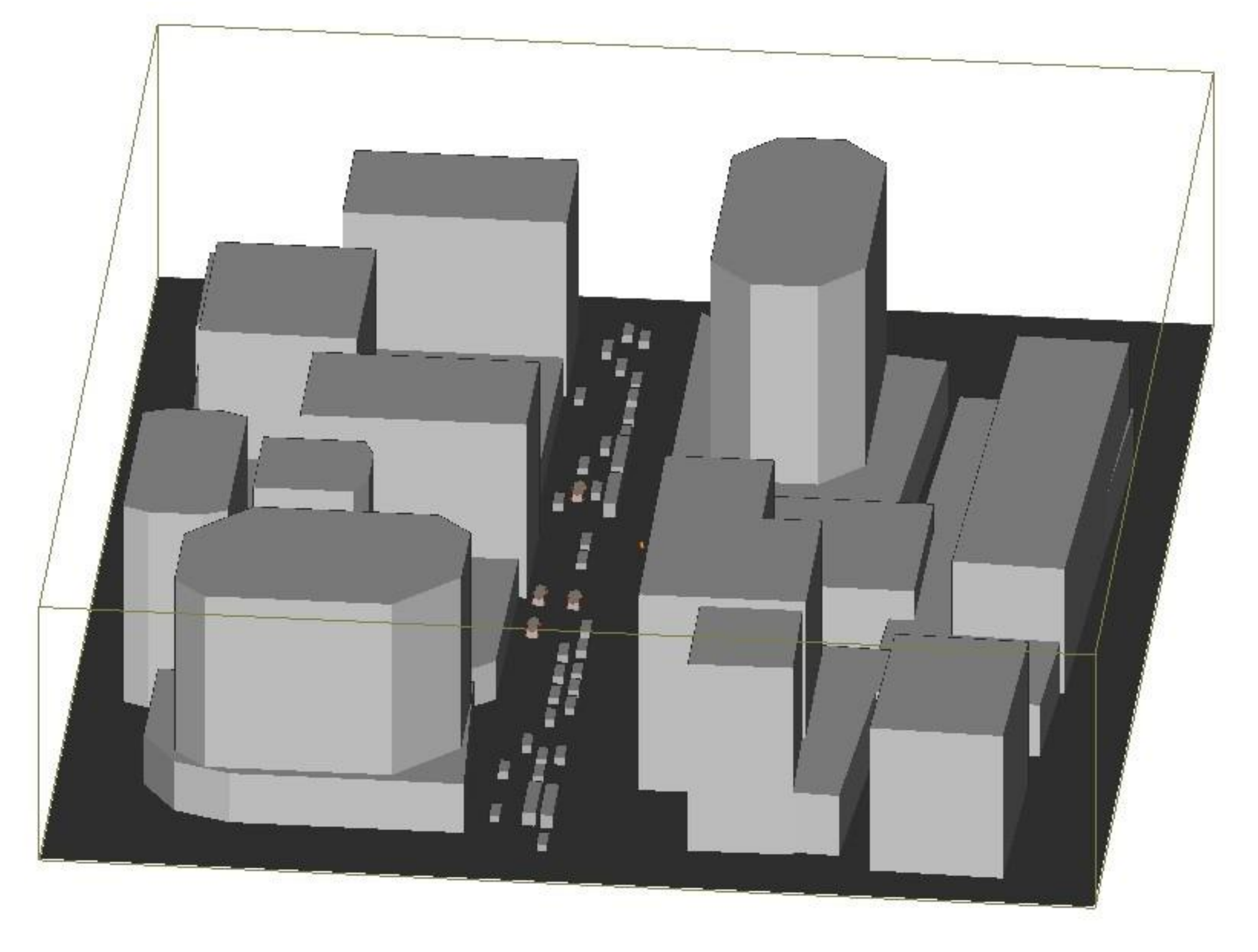}
\caption{The ray-tracing propagation environment. This models an urban roadway with 4 lanes.}
\label{fig:city}
\end{figure}

\begin{figure*}[h!]
\centering
\includegraphics[width=0.9\textwidth]{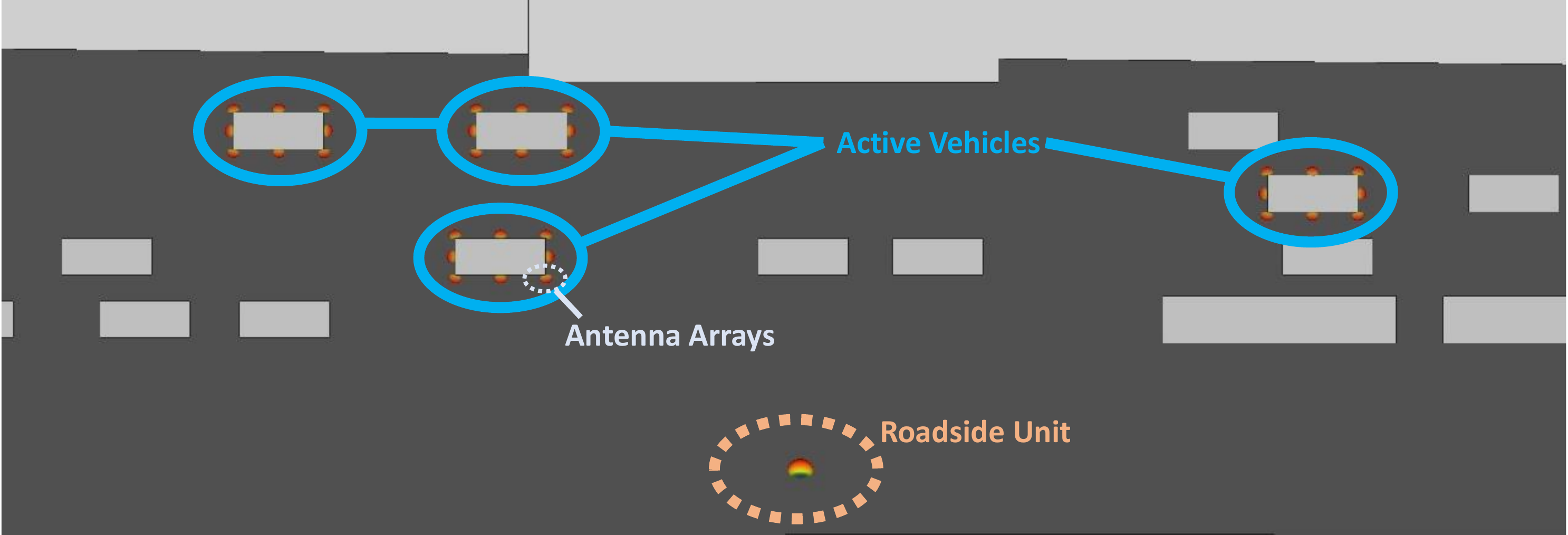}
\caption{An aerial view of the ray-tracing environment that shows the location of the RSU and multiple active vehicles equipped with both radar and communication transceivers.}
\label{fig:top_down}
\end{figure*}

\subsection{Simulation setup}\label{sec:simsetup} 
To generate our \ac{V2I} communication and radar channels, ray-tracing simulations were conducted in Wireless Insite \cite{WI}. The simulated communication channels operate at 73 GHz and the simulated radar channels operate at 76 GHz. The ray-tracing environment models an urban roadway, which is visualized in Fig.~\ref{fig:city}. In this environment, a mix of medium and large buildings are located along both sides of the roadway. For simulation purposes, the material of the buildings is assumed to be concrete with a relative permittivity of $5.31$, conductivity of $\SI{1.0509}{\siemens\per\meter}$ in the communication band at $\SI{73}{\giga\hertz}$, and a conductivity of $\SI{1.0858}{\siemens\per\meter}$ in the radar band at $\SI{76}{\giga\hertz}$ ~\cite[Table 3]{ITU2015Effects}. The surface of the roadway is assumed to be asphalt that has a relative permittivity of $3.18$, a conductivity of $\SI{0.4061}{\siemens\per\meter}$ at $\SI{73}{\giga\hertz}$, and a conductivity of $\SI{0.4227}{\siemens\per\meter}$ at $\SI{76}{\giga\hertz}$~\cite{Li1999Low}. \Ac{RMS} surface roughness is also modelled as $\SI{0.2}{\milli\meter}$ for concrete and $\SI{0.34}{\milli\meter}$ for asphalt~\cite[Table 1]{Li1999Low}. In Wireless Insite, diffuse scattering is parameterized by a scattering coefficient in the range $[0,1]$, which we select to be $0.4$ for concrete and $0.5$ for asphalt ~\cite{Remcom5G}. The fraction of diffuse reflections that experience cross-polarization is also parameterized with another coefficient in the range $[0,0.5]$, which we select to be $0.5$ for both concrete and asphalt. The material for the vehicles is assumed to be a perfect electric conductor metal.

The placement of the vehicles along the roadway is in accordance with option B for Urban scenarios as suggested by 3GPP ~\cite[6.1.2]{3GPP37885}. In this setup, we simulate 80\% of vehicles being cars of size $5\times2\times1.6\SI{}{\meter}$ and the remaining 20\% of vehicles being trucks of size $13\times2.6\times3\SI{}{\meter}$. We assume the the speed of vehicles is dependent upon the lane they are positioned in, and that the spacing between vehicles is exponentially distributed with a mean dependent upon the vehicle speed. The speed of vehicles in each lane are $60$, $50$, $25$, and $\SI{15}{\kilo\meter\per\hour}$, denoted $s_l$. For each vehicle, let $d_l \sim \text{Exp}(0.5/s_l)$. Then the distance from the previous vehicle is given by $\max(2,d_l)$~\cite[6.1.2]{3GPP37885}. Note that despite ascribing a velocity to each vehicle, each ray-tracing simulation operates on a static environment frozen in time. For each simulation, the type and placement of vehicles are generated randomly and independently according to the above distributions. After generating the vehicle placements, $M$ cars are selected within the area-of-coverage to be active vehicles equipped with radar and communication arrays. In our simulations, $M=4$ and the area-of-coverage is defined as the $60\SI{}{\meter}$ section of roadway centered around the \ac{RSU}.

The active cars are equipped with 4 communication arrays and 4 radar arrays. The communication arrays are placed at the front, sides, and rear of the car at a height of $1.6\SI{}{\meter}$. The radar arrays are placed at the 4 corners of the car with $\SI{10}{\degree}$ rotation toward the front or rear of the car and a height of $0.75 \SI{}{\meter}$. The antenna patterns for both the communication and radar antenna elements are chosen to have a half-power beamwidth of $\SI{120}{\degree}$. The arrays are assumed to be \acs{ULA} with inter-element spacing of half a wavelength.

\begin{figure}[h!]
\centering
\includegraphics[width=\columnwidth]{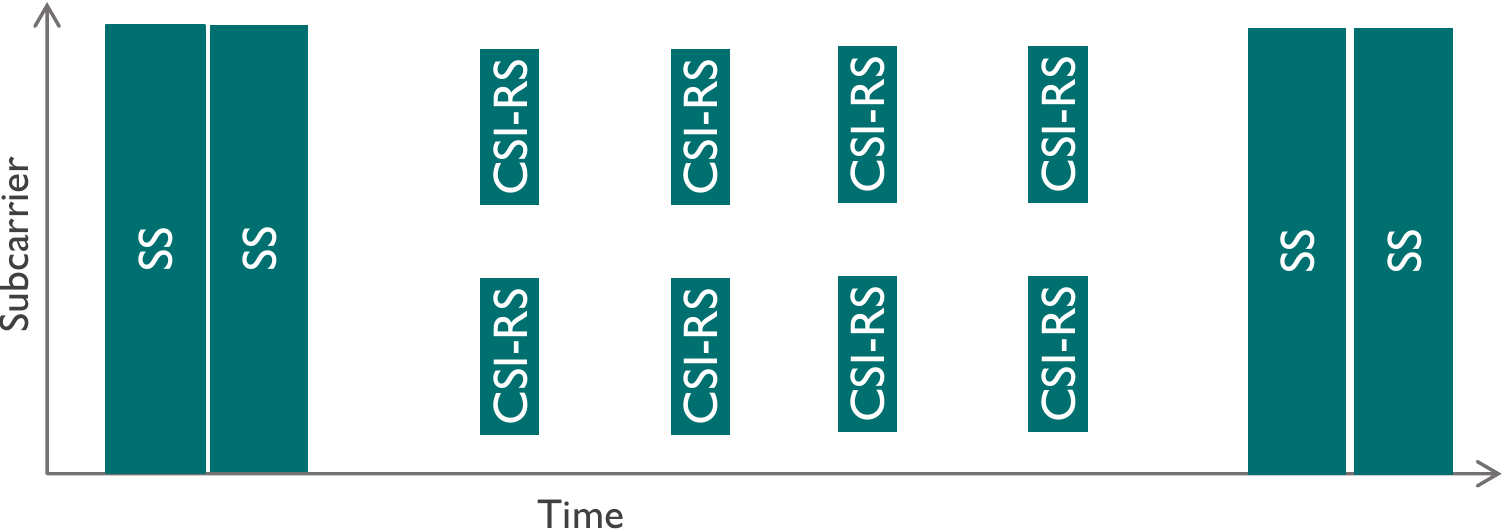}
\caption{The downlink-based protocol. SS blocks use 100\% of the subcarriers for training data. The CSI-RS blocks use 25\% of the subcarriers for training data to track the channel.}
\label{fig:dl}
\end{figure}

We simulate our communication link with a transmit power of $24~\dBm$. We assume rectangular pulse-shaping for simplicity. We use $K = 2048$ subcarriers with subcarrier spacing of $\SI{240}{\kilo\hertz}$. $D = 512$ time-domain channel taps are used to capture the delay spread of the ray-traced channels. Furthermore, a \ac{CP} of $D-1$ samples is included in each \ac{OFDM} symbol. At the \ac{RSU}, the system has $4$ 
RF-chains and $64$ antenna elements. At the vehicle, the system has $1$ RF-chain and $16$ antenna elements per array. The downlink protocol is visualized in Fig.~\ref{fig:dl}. This protocol is modeled after the standalone-downlink (SA-DL) scheme proposed in the beam management tutorial for 3GPP NR. In our protocol, synchronization (SS) blocks are transmitted every channel coherence time. Each SS block spans 100\% of the available subcarriers, uses 4 OFDM symbols, and supports up to $4$ simultaneous beams in accordance with the number of RF-chains. Tracking (CSI-RS) blocks are transmitted 4 times per coherence time. Each CSI-RS block spans 25\% of the available subcarriers, uses 1 OFDM symbol, and also supports up to 4 simultaneous beams. All other resources not occupied with SS or CSI-RS blocks are used to transmit data to the UE vehicles. We will explore 3 different protocol variants for initial access: exhaustive search, assisted narrow search, and assisted wide search. The exhaustive variant requires all possible beam configurations to be searched over when establishing a link. This corresponds to a typical approach unaided by out-of-band or radar information. The assisted variants use either the direct radar covariance estimates or any of the three neural network predictions. Since these predictions are aided, the search space of beam configurations can be drastically reduced compared to the exhaustive search. Table~\ref{table:search_s} lists the number of \ac{RSU} beams that are searched over for each variant, as well as the number of SS blocks required to search over that quantity of beams. This can be determined since each SS block supports 4 beams and the UE vehicle needs to search over 16 beams for every \ac{RSU}. The narrow search only uses 4 beams, while the wide search uses 12 beams. For too small of a search size, the best beam may not be selected. For too large of a search size, the training overhead may become too costly.

\begin{table}[!htbp]
\centering
\caption{Required search parameters for the three protocols.}
\begin{tabular}{lll}
\\ [-2ex]
\hline\hline \\ [-2ex]
\textbf{Protocol} & \textbf{Beam Search Size} & \textbf{SS Blocks} \\ [0.5ex] \hline \\ [-2ex]
Exhaustive Search                 & 64 & 256 \\
Assisted Narrow Search	          & 4  & 16 \\
Assisted Wide Search 			  & 12 & 48 \\
\hline
\end{tabular}
\label{table:search_s}
\end{table}

The FMCW radars are capable of transmitting at chirp rates in the range $[10,60] \SI{}{\mega\hertz\per\second}$. Each vehicle independently selects a chirp rate $\beta \sim \cU [10,60] \SI{}{\mega\hertz\per\second}$. Regardless of the chirp rate, each radar transmits an FMCW waveform with bandwidth of $B_r = \SI{1}{\giga\hertz}$ without downtime. The chirp period is defined as $T_p = \frac{B_r}{\beta}$. A random timing offset $\Delta{t} \sim \cU [0,T_p] \SI{}{\second}$ is also selected for each vehicle. Each vehicle uses this chirp rate and timing offset for all 4 of its radars, assuming that all 4 radars transmit in a synchronized manner. An additional random phase offset $\epsilon\sim\cU[0,2\pi]$ is added to each transmitted radar signal.

\subsection{Neural network training}

A learning dataset was generated using the true covariances of the communication channels and the estimated radar covariances after detection and filtering. 3000 independent environments were generated, each with 4 covariance pairs corresponding to the 4 active vehicles. This resulted in 12000 unique covariance training pairs. Of the 12000 pairs, 208 pairs were not detected from their radar transmissions and were discarded from the learning dataset, leaving 11792 properly detected pairs. This learning dataset was isolated and kept independent from the evaluation set used for the results shown later. The learning dataset was randomly sampled into a training dataset of 9434 entires and a validation dataset of 2358 entries. For each of the 3 neural networks, these covariances were pre-processed into APS's, dominant eigenvectors, and 1st columns of the covariance matrices.

The neural network was trained using the Adam optimizer. Early-stopping was used to halt training after a plateau of 16 training epochs and to restore the best weights that minimized the validation set. The learning-rate was also halved after plateaus of 6 training epochs, down to a minimum learning rate of $1e-6$. Training was run using Tensorflow and accelerated using a Nvidia GTX 1060 GPU.
\begin{figure}[h!]
\centering
\includegraphics[width=\columnwidth]{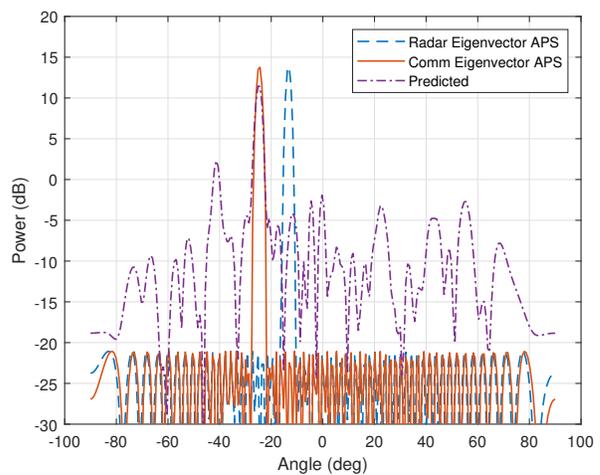}
\caption{An example of the trained network predicting an eigenvector input from the evaluation dataset. This example shows good alignment with the main beam of the communication APS and the predicted eigenvector's APS.}
\label{fig:predicted}
\end{figure}

\begin{figure}[h!]
\centering
\includegraphics[width=\columnwidth]{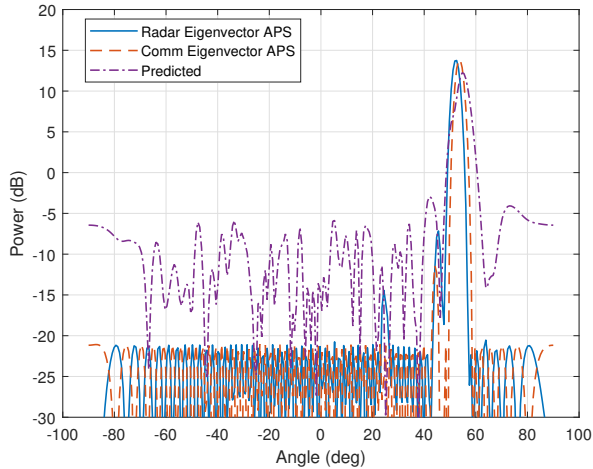}
\caption{A second example of the trained network predicting an eigenvector input from the evaluation dataset. This example shows prediction where there is still some angular error between the peak of the predicted APS and the communication APS.}
\label{fig:predicted2}
\end{figure}

\subsection{MmWave link configuration}\label{sec:mmW} 

Using the parameters above, we will now simulate the communication link in a Monte Carlo fashion to characterize its performance and compare the three variants described. 1000 independent environments were generated. For each environment, one active vehicle was selected at random to be in initial access, while the remaining three vehicles were selected to be in tracking.

As mentioned in Section \ref{sec:simsetup}, three beam training protocols were evaluated: exhaustive search, assisted narrow search, and assisted wide search. Each protocol defines a search space of beams which will evaluated during the SS transmissions and requires a different number of SS blocks to be sent. Assume the \ac{RSU} uses the same 2-bit DFT codebook for its $N_{\text{RSU}}=64$ beams, and that the UE uses the same 2-bit DFT codebook for its $N_{\text{V}}=16$ beams. Let $\mathcal{Q}(.)$ define the 2-bit quantization function. Then the \ac{RSU} codebook $\bC_{\text{RSU}}$ is defined as
\begin{multline}
    [\bC_{\text{RSU}}]_i = \mathcal{Q}\left(\frac{1}{N_{\text{RSU}}}\baRSU\left( \arcsin{\left(\frac{2i-N_{\text{RSU}}-1}{N_{\text{RSU}}}\right)} \right)\right) \\ \forall i \in [N_{\text{RSU}}].
\end{multline}
Similarly, the UE codebook $\bC_{\text{V}}$ is defined as
\begin{multline}
    [\bC_{\text{V}}]_i = \mathcal{Q}\left(\frac{1}{N_{\text{V}}}\baRSU\left( \arcsin{\left(\frac{2i-N_{\text{V}}-1}{N_{\text{V}}}\right)} \right)\right) \\ \forall i \in [N_{\text{V}}].
\end{multline}

Let the beam search spaces for vehicle $i$ be defined as a set $S_{\text{RSU},i}$ and $S_{\text{V},i}$. Then the best RF-precoder $\mathbf{f}_{i}$ and RF-combiner $\bw_{i}$ are selected as
\begin{equation}
    \{\mathbf{f}_{i},\bw_{i}\} = \arg \max_{\mathbf{f}_{i} \in S_{\text{RSU}},\bw_{i} \in S_{\text{V}}} \sum_{k=1}^{K} \log_2 \left( 1 + (\bw_{i}^\ast \bH_{i}[k] \mathbf{f}_{i})^2 \right).
\end{equation}
We assume that the vehicles in tracking always select the best pair of RF-precoders and RF-combiners, so for those vehicles $S_{\text{RSU},i} = [N_{\text{RSU}}]$ and $S_{\text{V},i} = [N_{\text{V}}]$. Let the selected RF-precoders and RF-combiners be stacked into matrices such that $[\bFRF]_i = \mathbf{f}_{i}$ and $[\bWRF]_i = \bw_{i}$. If we assume that $\bFBB[k]$ and $\bWBB[k]$ are diagonal matrices, then the signal power and interference power for the stream to UE $i$ at subcarrier $k$ can be defined as 
\begin{equation}
    P_{\text{sig},i}[k] = ([\bWRF]_{i}^\ast \bH_{i}[k] [\bFRF]_{i})^2,
\end{equation}
and
\begin{equation}
    P_{\text{int},i}[k] = \sum_{l \neq i} ([\bWRF]_{l}^\ast \bH_{i}[k] [\bFRF]_{l})^2.
\end{equation}
Recall that we transmit each stream with a power of $P_t = 24~\dBm$. With equal power allocation across all subcarriers, each subcarrier in each stream has a transmit power of $P_t = -9.1~\dBm$. Assume a thermal noise of $N_0 = -174~\dBm/\SI{}{\hertz}$, a system noise factor of $10 \SI{}{\decibel}$, and a subcarrier spacing of $B_{\text{sc}} = \SI{240}{\mega\hertz}$. Then the noise power per subcarrier is $P_n = -110.2~\dBm$. The signal-to-interference-noise ratio (SINR) can be defined as
\begin{equation}
    \text{SINR}_{i}[k] = \frac{P_{\text{sig},i}[k] P_t}{P_{\text{int},i}[k] P_t + P_n}.
\end{equation}

After computing the SINR, the spectral efficiency can be defined as
\begin{equation}
    s_i = \sum_{k=1}^{K} \log_2 \left( 1 + \text{SINR}_{i}[k] \right).
\end{equation}
Let $T_{\text{train}}$ be the effective time spent transmitting training data from the SS and CSI-RS blocks. ``Effective" refers to an average over all subcarriers. Let $N_{\text{SS}}$ be the number of SS blocks and $N_{\text{CSI-RS}}$ be the number of CSI-RS blocks sent in the transmitted frame. Also let $\nu$ denote the fraction of subcarriers used by CSI-RS blocks. The effective training time is then computed as
\begin{equation}
    T_{\text{train}} = T_{\text{sym}} \frac{N_{\text{SS}} N_{\text{sym-per-SS}}  + \nu{} N_{\text{CSI-RS}} N_{\text{sym-per-CSI-RS}}}{N_{\text{beams}}}.
\end{equation}
 Now define $T_{\text{coh}}$ as the channel coherence time and assume that the communication system repeats its training protocol every $T_{\text{coh}}$ seconds. We can then compute the effective rate $R_i$ for each link as
\begin{equation}
    R_i = (1-\frac{T_{\text{train}}}{T_{\text{coh}}}) B_{\text{sc}} s_i.
\end{equation}
The sum-rate is then defined as 
\begin{equation}
    R_{\Sigma} = \sum_{i} R_i.
\end{equation}
\begin{figure}[h!]
\centering
\includegraphics[trim={1cm 0cm 2.5cm 1.5cm},clip,width=\columnwidth]{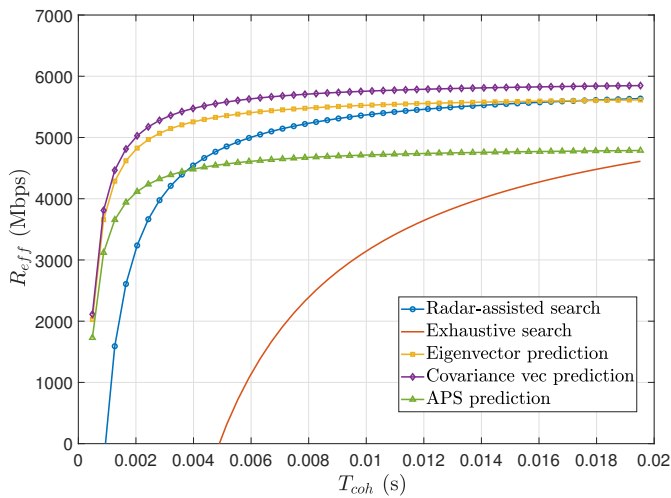}
\caption{The Monte Carlo sum rate results over all 4 users using an exhaustive search and all versions of the assisted narrow search.}
\label{fig:4_rate}
\end{figure}

\begin{figure}[h!]
\centering
\includegraphics[trim={1cm 0cm 2.5cm 1.5cm},clip,width=\columnwidth]{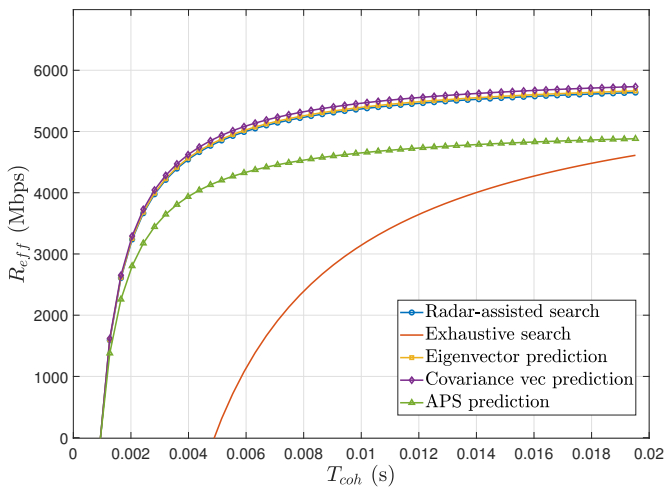}
\caption{The Monte Carlo sum rate results over all 4 users using an exhaustive search and all versions of the assisted wide search.}
\label{fig:12_rate}
\end{figure}

The sum-rate results are plotted in Fig.~\ref{fig:4_rate} for the assisted narrow search and Fig.~\ref{fig:12_rate} for the assisted wide search. Both plots include the exhaustive search results for comparison. Due to the large training overhead of the exhaustive search, the exhaustive strategy achieves poor sum-rates for short coherence times. Below \SI{0.005}{\second}, the exhaustive strategy requires more training symbols than what can be fit within one coherence interval, resulting in no data transmission. Comparatively, the assisted strategies allow the links to be established for even short coherence intervals. In Fig.~\ref{fig:4_rate}, the assisted search based on covariance vector prediction achieves the highest rate for the entire range of coherence times, with the assisted search based on the eigenvector prediction slightly below. The pure radar-assisted search without learning based radar-to-communication mapping has similar performance at long coherence times, but the performance declines much faster at shorter times. The APS prediction neural network provides the lowest sum-rates of the assisted strategies at longer coherence times, yet outperforms the radar-assisted search at very short coherence times. All of the assisted strategies show significant rate improvements over the exhaustive search, which is infeasible for short coherence times due to the large training overhead required. In Fig.~\ref{fig:12_rate}, all the assisted methods use slightly more training overhead to reduce the likelihood that the optimal beam is missed. As a result, the difference in sum-rates between assisted strategies based on covariance vector prediction, eigenvector prediction, and pure radar-assisted is reduced.

\subsection{Outage and Detection Errors}

Our system can experience two main failures. The first is a failure to detect an FMCW signal that was transmitted. This will be called the probability of missed detection $P_m$. The second is failure to achieve a high enough SNR in the communication link to transmit at a meaningful data rate. This will be called the probability of outage $P_o$. These metrics are evaluated on the vehicle in initial access for each simulation instance. Of the 1000 instances each with a single vehicle in initial access, the FMCW transmissions were not detected in 15 instances giving $P_m = 0.015$. Of the remaining 985 instances, the sum rate results were computed. Let us define the minimum supported rate as $R_{\text{min}}$ and the rate of the vehicle in intial access in instance $i$ out of $N_i$ total instances as $R_i$. Then the probability of outage is defined as

\begin{equation}
    P_o = \frac{1}{N_i}\sum_{i=1}^{N_i} \mathbbm{1}(R_i < R_{\text{min}}).
\end{equation}

\begin{figure}[h!]
\centering
\includegraphics[trim={1cm 0cm 2.5cm 1.5cm},clip,width=\columnwidth]{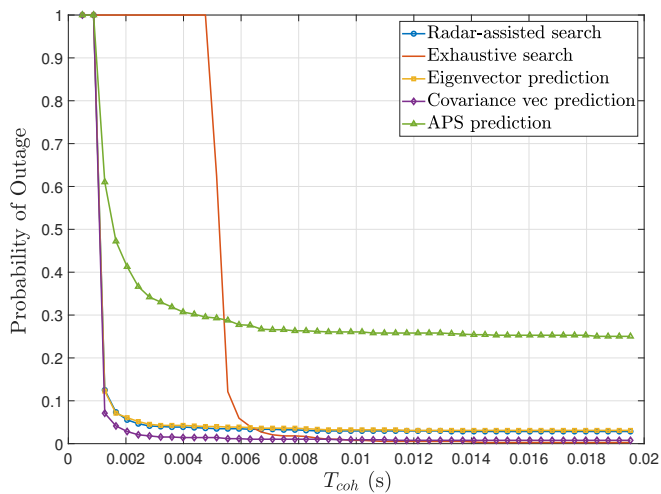}
\caption{The LOS probability of outage for the vehicle in initial access with an outage rate of $R_{\text{min}} = \SI{100}{\mega\bitspersecond}$.}
\label{fig:po_los}
\end{figure}

\begin{figure}[h!]
\centering
\includegraphics[trim={1cm 0cm 2.5cm 1.5cm},clip,width=\columnwidth]{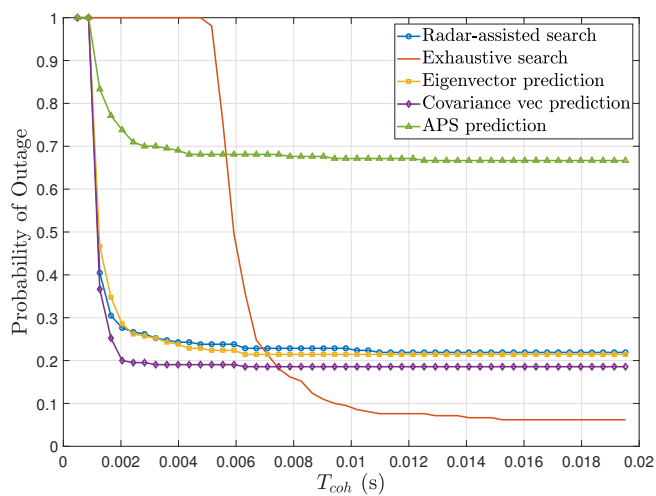}
\caption{The NLOS probability of outage for the vehicle in initial access with an outage rate of $R_{\text{min}} = \SI{100}{\mega\bitspersecond}$.}
\label{fig:po_nlos}
\end{figure}

Set $R_{\text{min}} = \SI{100}{\mega\bitspersecond}$. Of the 985 instances with correct detection, 775 were classified as LOS and 210 were classified as NLOS. The assisted wide search (12 beams) was used for these results. The LOS probability of outage is shown in Fig.~\ref{fig:po_los}, and the NLOS probability of outage is shown in Fig.~\ref{fig:po_nlos}. As expected, the NLOS cases tend to have a higher probability of outage for all beam training methods. Below coherence times of $T_{coh} = \SI{5}{\milli\second}$, the exhaustive search requires too many spectral resources to complete before the channel is incoherent, resulting in a probability of outage of $1$. In both the LOS and NLOS cases, the covariance vector prediction neural network yields a reduction in the probability of outage compared to all other methods for coherence times below approximately \SI{6}{\milli\second}. Eigenvector prediction and pure radar-assisted strategies have roughly similar performance, while APS prediction has a notably higher outage probability in the same region. This demonstrates that our assisted strategies provide clear benefits over an exhaustive search in both LOS and NLOS propagation environments at short coherence times. Furthermore, learning-based assisted methods, especially the covariance vector prediction, can provide even further benefits compared to a pure radar-assisted search.

\section{Conclusion and future work}\label{sec:conc}

In this work, we extended the application of radar-assisted beam training to multi-user communications. We designed a processing chain to estimate the individual spatial covariances of multiple interfering FMCW signals measured at a passive antenna array, and trained a neural network to predict selected features of the communication spatial covariances based on the estimated radar covariances. The proposed approaches were evaluated against a traditional exhaustive beam search strategy and an approach that used the APS of the estimated radar covariance without any further refinement based on a neural network. The features used for neural network prediction include the APS, the dominant eigenvector, and the covariance vector. Ray tracing software was used to generate mmWave radar and communication channels to create a training set for optimizing the neural network, and an evaluation dataset for comparing the beam training methods. The proposed assisted methods show drastic increases in the sum-rate compared to the exhaustive search. Our study additionally showed that the assisted search based on the learned  covariance vector provides higher sum-rates and lower probability of outage than all other assisted methods, in both LOS and NLOS environments.

These results show that out-of-band spatial information from passive radars can be used for multi-user systems when special processing is implemented to estimate signal parameters and filter out interference. In addition, the intricate differences between radar spatial characteristics and the communication channel spatial characteristics can be learned through deep learning, yielding more accurate channel predictions even in NLOS channels. This improved accuracy allows the system to reduce the size of its beam search, significantly reducing overhead and increasing the data rate.

Future work should explore the detection of automotive radars beyond FMCW, including OFDM and PMCW radar. On the communications side, more intelligent power allocation and subcarrier allocation should also be explored. The channels generated from the ray tracing simulations were also static channels. While it may be computationally difficult, simulating dynamic channels over time would allow for even more realistic analysis of the real world performance of the proposed beam training approach.

\bibliographystyle{IEEEtran}
\bibliography{refs}

\begin{thebibliography}{10}
\providecommand{\url}[1]{#1}
\csname url@samestyle\endcsname
\providecommand{\newblock}{\relax}
\providecommand{\bibinfo}[2]{#2}
\providecommand{\BIBentrySTDinterwordspacing}{\spaceskip=0pt\relax}
\providecommand{\BIBentryALTinterwordstretchfactor}{4}
\providecommand{\BIBentryALTinterwordspacing}{\spaceskip=\fontdimen2\font plus
\BIBentryALTinterwordstretchfactor\fontdimen3\font minus
  \fontdimen4\font\relax}
\providecommand{\BIBforeignlanguage}[2]{{%
\expandafter\ifx\csname l@#1\endcsname\relax
\typeout{** WARNING: IEEEtran.bst: No hyphenation pattern has been}%
\typeout{** loaded for the language `#1'. Using the pattern for}%
\typeout{** the default language instead.}%
\else
\language=\csname l@#1\endcsname
\fi
#2}}
\providecommand{\BIBdecl}{\relax}
\BIBdecl

\bibitem{Alland2019Interference}
S.~Alland, W.~Stark, M.~Ali, and M.~Hegde, ``Interference in automotive radar
  systems: characteristics, mitigation techniques, and current and future
  research,'' \emph{IEEE Signal Processing Magazine}, vol.~36, no.~5, pp.
  45--59, 2019.

\bibitem{Ali2020Passive}
A.~Ali, N.~Gonz{\'a}lez-Prelcic, and A.~Ghosh, ``Passive radar at the roadside
  unit to configure millimeter wave vehicle-to-infrastructure links,''
  \emph{IEEE Transactions on Vehicular Technology}, vol.~69, no.~12, pp.
  14\,903--14\,917, 2020.

\bibitem{Prelcic2017}
N.~Gonz{\'a}lez-Prelcic, A.~Ali, V.~Va, and R.~W. Heath, ``Millimeter-wave
  communication with out-of-band information,'' \emph{IEEE Communications
  Magazine}, vol.~55, no.~12, pp. 140--146, 2017.

\bibitem{Hashemi2018OOB}
M.~Hashemi, C.~E. Koksal, and N.~B. Shroff, ``Out-of-band millimeter wave
  beamforming and communications to achieve low latency and high energy
  efficiency in {5G} systems,'' \emph{IEEE Transactions on Communications},
  vol.~66, no.~2, pp. 875--888, 2018.

\bibitem{Ali2018OOB}
A.~Ali, N.~Gonz{\'a}lez-Prelcic, and R.~W. Heath, ``Millimeter wave
  beam-selection using out-of-band spatial information,'' \emph{IEEE
  Transactions on Wireless Communications}, vol.~17, no.~2, pp. 1038--1052,
  2018.

\bibitem{Ali2019Covariance}
------, ``Spatial covariance estimation for millimeter wave hybrid systems
  using out-of-band information,'' \emph{IEEE Transactions on Wireless
  Communications}, vol.~18, no.~12, pp. 5471--5485, 2019.

\bibitem{Prelcic2016Radar}
N.~Gonz{\'a}lez-Prelcic, R.~Méndez-Rial, and R.~W. Heath, ``Radar aided beam
  alignment in {mmWave} {V2I} communications supporting antenna diversity,'' in
  \emph{2016 Information Theory and Applications Workshop (ITA)}, 2016, pp.
  1--7.

\bibitem{Liu2020Radar}
F.~Liu, W.~Yuan, C.~Masouros, and J.~Yuan, ``Radar-assisted predictive
  beamforming for vehicular links: communication served by sensing,''
  \emph{IEEE Transactions on Wireless Communications}, vol.~19, no.~11, pp.
  7704--7719, 2020.

\bibitem{Klautau2019Lidar}
A.~Klautau, N.~Gonz{\'a}lez-Prelcic, and R.~W. Heath, ``{LIDAR} data for deep
  learning-based {mmWave} beam-selection,'' \emph{IEEE Wireless Communications
  Letters}, vol.~8, no.~3, pp. 909--912, 2019.

\bibitem{Brambilla2019IMU}
M.~Brambilla, M.~Nicoli, S.~Savaresi, and U.~Spagnolini, ``Inertial sensor
  aided {mmWave} beam tracking to support cooperative autonomous driving,'' in
  \emph{2019 IEEE International Conference on Communications Workshops (ICC
  Workshops)}, 2019, pp. 1--6.

\bibitem{Kela2016Location}
P.~Kela, M.~Costa, J.~Turkka, M.~Koivisto, J.~Werner, A.~Hakkarainen,
  M.~Valkama, R.~Jantti, and K.~Leppanen, ``Location based beamforming in {5G}
  ultra-dense networks,'' in \emph{2016 IEEE 84th Vehicular Technology
  Conference (VTC-Fall)}, 2016, pp. 1--7.

\bibitem{Miao2019UAV}
W.~Miao, C.~Luo, G.~Min, L.~Wu, T.~Zhao, and Y.~Mi, ``Position-based
  beamforming design for {UAV} communications in {LTE} networks,'' in \emph{ICC
  2019 - 2019 IEEE International Conference on Communications (ICC)}, 2019, pp.
  1--6.

\bibitem{VaEtAlPositionaidedMillimeterWaveV2I2017}
V.~Va, T.~Shimizu, G.~Bansal, and R.~W. Heath, ``Position-aided millimeter wave
  {V2I} beam alignment: a learning-to-rank approach,'' in \emph{2017 {{IEEE}}
  28th {{Annual International Symposium}} on {{Personal}}, {{Indoor}}, and
  {{Mobile Radio Communications}} ({{PIMRC}})}, Oct. 2017, pp. 1--5.

\bibitem{V.VaEtAlInverseMultipathFingerprintingMillimeter2018}
{V. Va}, {J. Choi}, {T. Shimizu}, {G. Bansal}, and R.~W. Heath, ``Inverse
  multipath fingerprinting for millimeter wave {V2I} beam alignment,''
  \emph{IEEE Transactions on Vehicular Technology}, vol.~67, no.~5, pp.
  4042--4058, May 2018.

\bibitem{V.VaEtAlOnlineLearningPositionAidedMillimeter2019}
{V. Va}, {T. Shimizu}, {G. Bansal}, and R.~W. Heath, ``Online learning for
  position-aided millimeter wave beam training,'' \emph{IEEE Access}, vol.~7,
  pp. 30\,507--30\,526, 2019.

\bibitem{SatyanarayanaEtAlDeepLearningAidedFingerprintbased2019}
K.~Satyanarayana, M.~{El-Hajjar}, A.~A.~M. Mourad, and L.~Hanzo, ``Deep
  learning aided fingerprint-based beam alignment for {{mmWave}} vehicular
  communication,'' \emph{IEEE Transactions on Vehicular Technology}, vol.~68,
  no.~11, pp. 10\,858--10\,871, Nov. 2019.

\bibitem{Y.WangEtAlMmWaveVehicularBeamSelection2019}
{Y. Wang}, {A. Klautau}, {M. Ribero}, {A. C. K. Soong}, and R.~W. Heath,
  ``{MmWave} vehicular beam selection with situational awareness using machine
  learning,'' \emph{IEEE Access}, vol.~7, pp. 87\,479--87\,493, 2019.

\bibitem{SimEtAlOnlineContextawareMachineLearning2018a}
G.~H. Sim, S.~Klos, A.~Asadi, A.~Klein, and M.~Hollick, ``An online
  context-aware machine learning algorithm for {{5G mmWave}} vehicular
  communications,'' \emph{IEEE/ACM Transactions on Networking}, vol.~26, no.~6,
  pp. 2487--2500, 2018.

\bibitem{Javier2018ICC}
J.~Rodriguez-Fernandez, N.~Gonz{\'a}lez-Prelcic, and R.~W. Heath,
  ``Position-aided compressive channel estimation and tracking for millimeter
  wave multi-user {MIMO} air-to-air communications,'' in \emph{2018 IEEE
  International Conference on Communications Workshops (ICC Workshops)}, 2018,
  pp. 1--6.

\bibitem{Javier2019SPAWC}
J.~Rodriguez-Fernandez, N.~Gonz{\'a}lez-Prelcic, I.~Pamplona-Trindade, and
  A.~Klautau, ``Position-aided compressive channel estimation and tracking for
  millimeter wave multi-user {MIMO} air-to-ground communications,'' in
  \emph{2019 IEEE 20th International Workshop on Signal Processing Advances in
  Wireless Communications (SPAWC)}, 2019, pp. 1--5.

\bibitem{mashhadi2021federated}
\BIBentryALTinterwordspacing
M.~B. Mashhadi, M.~Jankowski, T.~Tung, S.~Kobus, and D.~G{\"{u}}nd{\"{u}}z,
  ``Federated mmwave beam selection utilizing {LIDAR} data,'' \emph{CoRR}, vol.
  abs/2102.02802, 2021. [Online]. Available:
  \url{https://arxiv.org/abs/2102.02802}
\BIBentrySTDinterwordspacing

\bibitem{zecchin2021lidar}
\BIBentryALTinterwordspacing
M.~Zecchin, M.~B. Mashhadi, M.~Jankowski, D.~G{\"{u}}nd{\"{u}}z, M.~Kountouris,
  and D.~Gesbert, ``A novel look at lidar-aided data-driven mmwave beam
  selection,'' \emph{CoRR}, vol. abs/2104.14579, 2021. [Online]. Available:
  \url{https://arxiv.org/abs/2104.14579}
\BIBentrySTDinterwordspacing

\bibitem{AliRadarConf2019}
A.~Ali, N.~Gonz\'{a}lez-Prelcic, and A.~Ghosh, ``{Millimeter wave V2I
  beam-training using base-station mounted radar},'' in \emph{2019 IEEE Radar
  Conference (RadarConf)}, 2019, pp. 1--5.

\bibitem{Graff2019}
A.~Graff, A.~Ali, and N.~Gonz\'{a}lez-Prelcic, ``Measuring radar and
  communication congruence at millimeter wave frequencies,'' in \emph{53rd
  Asilomar Conference on Signals, Systems, and Computers}, 2019, pp. 925--929.

\bibitem{ChenGlobecom2021}
Y.~Chen, A.~Graff, N.~Gonz{\'a}lez-Prelcic, and T.~Shimizu, ``Radar aided
  {mmWave} vehicle-to-infrastructure link configuration using deep learning,''
  in \emph{IEEE Global Communications Conference}, 2021.

\bibitem{3GPP37885}
\BIBentryALTinterwordspacing
3GPP, ``Study on evaluation methodology of new vehicle-to-everything {(V2X)}
  use cases for {LTE} and {NR},'' {3rd Generation Partnership Project (3GPP)},
  TR 37.885, Sep. 2018, version 15.1.0. [Online]. Available:
  \url{http://www.3gpp.org/DynaReport/37885.htm}
\BIBentrySTDinterwordspacing

\bibitem{Alkhateeb2016Frequency}
A.~Alkhateeb and R.~W. Heath~Jr., ``{Frequency selective hybrid precoding for
  limited feedback millimeter wave systems},'' vol.~64, no.~5, pp. 1801--1818,
  May 2016.

\bibitem{Bjornson2009Exploiting}
E.~Bjornson, D.~Hammarwall, and B.~Ottersten, ``{Exploiting quantized channel
  norm feedback through conditional statistics in arbitrarily correlated MIMO
  systems},'' vol.~57, no.~10, pp. 4027--4041, 2009.

\bibitem{Katkovnik2002High}
V.~Katkovnik, M.-S. Lee, and Y.-H. Kim, ``{High-resolution signal processing
  for a switch antenna array FMCW radar with a single channel receiver},''
  2002, pp. 543--547.

\bibitem{Cumplido2004}
R.~Cumplido, C.~Torres, and S.~Lopez, ``On the implementation of an efficient
  {FPGA}-based {CFAR} processor for target detection,'' in \emph{(ICEEE). 1st
  International Conference on Electrical and Electronics Engineering, 2004.},
  2004, pp. 214--218.

\bibitem{kalman1992tanh}
B.~L. Kalman and S.~C. Kwasny, ``Why tanh: choosing a sigmoidal function,'' in
  \emph{[Proceedings 1992] IJCNN International Joint Conference on Neural
  Networks}, vol.~4.\hskip 1em plus 0.5em minus 0.4em\relax IEEE, 1992, pp.
  578--581.

\bibitem{WI}
``{Wireless Insite},'' \url{http://www.remcom.com/wireless-insite}.

\bibitem{ITU2015Effects}
I.~T.~U. (ITU), ``{Effects of building materials and structures on radiowave
  propagation above about 100 MHz},'' Tech. Rep. ITU-R P.2040-1, Jul. 2015.

\bibitem{Li1999Low}
E.~S. Li and K.~Sarabandi, ``{Low grazing incidence millimeter-wave scattering
  models and measurements for various road surfaces},'' vol.~47, no.~5, pp.
  851--861, 1999.

\bibitem{Remcom5G}
Remcom, ``{5G} mmwave channel modeling with diffuse scattering in an office
  environment.''

\end{thebibliography}
\end{document}